\documentclass{aa}
\usepackage{natbib,epsfig,amsmath,txfonts,graphicx}

\def\lam {$\lambda$}

\def\HI  {H{\sc i}}
\def\HII {H{\sc ii}}
\def\HeI {He{\sc i}}
\def\HeII{He{\sc ii}}

\def\NIII{N{\sc iii}}
\def\NV{N{\sc v}}
\def\CIII{C{\sc iii}}
\def\CIV{C{\sc iv}}
\def\teff{$T_{\rm eff}$}
\def\g   {$\log g$}
\def\Y   {$Y_{\rm He}$}

\def\Rsun {$R_{\odot}$}
\def\Rsune {R_{\odot}}

\def\Msune {{\rm M}_{\odot}}

\def\Rstar{$R_{\ast}$}

\def\Rsuna{$\rm{R_{\odot}}$} 
 
\def\Msuna{$\rm{M_{\odot}}$} 

\def\Mdot{${\dot M}$}

\def\MV {M$_{\rm v}$}
\def\Vr {$V_{\rm r}${\thinspace}sin{\thinspace}$i$}

\def\vsini {v \sin i} 

\def\Vi {$v_{\rm \infty}$}

\def\Hg {H$_{\rm \gamma}$}

\def\Ha {H$_{\rm \alpha}$}

\def\Bra{Br$_{\rm \alpha}$}
\def\Brg{Br$_{\rm \gamma}$}
\def\Pfg{Pf$_{\rm \gamma}$}
\def\Pfb{Pf$_{\rm \beta}$}
\def\kms {km~s$^{-1}$}
\def\Mdu{$\cdot 10^{-6} {\rm M_{\odot}/yr}$} 

\def\rarrow{$\rightarrow$}

\def\vturb{$v_{\rm turb}$}

\def\beq{\begin{equation}}
\def\eeq{\end{equation}}
\def\beqa{\begin{eqnarray}}
\def\eeqa{\end{eqnarray}}

\def \Rstare {R_\star}
\def \Teffe {T_{\rm eff}}
\def \Te {T_{\rm e}}
\def \vinfe {v_\infty}
\def \Mdote {\dot M}

\def \taur {\tau_{\rm Ross}}

\begin{document}

\title{Quantitative H and K band spectroscopy of Galactic OB-stars at medium resolution}

\subtitle{}

\author{T. Repolust\inst{1}, J. Puls\inst{1}, M.M. Hanson\inst{2,3}, 
R.-P. Kudritzki\inst{4}, M.R. Mokiem\inst{5}}

\offprints{T. Repolust}

\institute{Universit\"ats-Sternwarte M\"unchen, Scheinerstr. 1, 
D-81679 M\"unchen, Germany\\
\email{repo@usm.uni-muenchen.de, uh101aw@usm.uni-muenchen.de}
\and
Department of Physics, The University of Cincinnati, Cincinnati, OH 45221-0011\\
\email{hanson@physics.uc.edu}
\and
Visiting Astronomer, Subaru Observatory at Mauna Kea, Hawaii
\and
Institute for Astronomy, University of Hawaii at Manoa, 2680 Woodlawn Drive,
Honolulu, HI 96822\\
\email{kud@ifa.hawaii.edu}
\and
Astronomical Institute ``Anton Pannekoek'', Kruislaan 403, NL-1098 SJ
Amsterdam\\
\email{mokiem@science.uva.nl}
}

\date{Received; accepted }

\abstract{In this paper we have analyzed 25 Galactic O and early
B-stars by means of {\it H} and {\it K} band spectroscopy, with the primary
goal to investigate to what extent a lone near-IR spectroscopy is able to
recover stellar and wind parameters derived in the optical. Most of the
spectra have been taken with {\sc subaru-ircs}, at an intermediate
resolution of 12,000, and with a very high S/N, mostly on the order of 200 or
better. In order to synthesize the strategic H/He lines, we have used our
recent, line-blanketed version of {\sc fastwind} (\citealt{puls05}). In
total, seven lines have been investigated, where for two stars we could make
additional use of the \HeI2.05 singlet which has been observed with {\sc
irtf-cshell}. Apart from \Brg\ and \HeII2.18, the other lines are
predominately formed in the stellar photosphere, and thus remain fairly
uncontaminated from more complex physical processes, particularly clumping.

First we investigated the predicted behaviour of the strategic lines. In
contradiction to what one expects from the optical in the O-star regime, 
almost all photospheric H/\HeI/\HeII\ {\it H/K} band lines {\it become
stronger if the gravity decreases}. Concerning H and \HeII, this finding is
related to the behaviour of Stark broadening as a function of electron
density, which in the line cores is different for members of lower (optical)
and higher (IR) series. Regarding \HeI, the predicted behaviour is due to
some subtle NLTE effects resulting in a stronger overpopulation of the lower
level when the gravity decreases.

We have compared our calculations with results from the alternative NLTE
model atmosphere code {\sc cmfgen} (\citealt{hil98}). In most cases, we
found reasonable or nearly perfect agreement. Only the \HeI2.05 singlet for
mid O-types suffers from some discrepancy, analogous with
findings for the optical \HeI\ singlets.

For most of our objects, we obtained good fits, except for the line cores of
\Brg\ in early O-stars with significant mass-loss. Whereas the observations
show \Brg\ mostly as rather symmetric emission lines, the models predict a
P~Cygni type profile with strong absorption. This discrepancy (which also
appears in lines synthesized by {\sc cmfgen}) might be an indirect effect of
clumping.

After having derived the stellar and wind parameters from the IR, we have
compared them to results from previous optical analyses. Overall, the IR
results coincide in most cases with the optical ones within the typical
errors usually quoted for the corresponding parameters, i.e, an uncertainty
in \teff\ of 5\%, in \g\ of 0.1 dex and in \Mdot\ of 0.2 dex, with lower
errors at higher wind densities. Outliers above the 1-$\sigma$ level where
found in four cases with respect to \g\ and in two cases for \Mdot. 

}

\titlerunning{Quantitative H and K band spectroscopy of Galactic OB-stars}
\authorrunning{T. Repolust et al.}

\maketitle

\section{Introduction}
\label{intro}

Although rare by number, massive stars dominate the life cycle of gas and
dust in star forming regions. They are responsible for the chemical enrichment
of the ISM, which in turn has a significant impact on the chemical evolution
of the parent galaxy. The main reason for this is that due to their large
masses, each physical stage evolves on much shorter timescales and more
violently, compared to low-mass stars, which provides a very efficient
recycling of elements. Moreover, the large amount of momentum and
energy input of these objects into the ISM controls the dynamical evolution
of the ISM and, in turn, the evolution of the parent galaxy (e.g., 
\citealt{LeithHeck95, ST01, Oey03}).

Presently, high mass star formation is still poorly understood. By
their nature, all star forming regions are found buried in molecular gas and
dust, allowing little or no light to escape at optical wavelengths. But,
unlike low mass stars which may eventually become optically visible while
still contracting to the main sequence, the short contraction timescale of
high mass stars keeps them deeply embedded throughout the entire formation
process. Compared to the optical, the dust and gas surrounding young massive
stars become more transparent in the infrared (IR) regime. Observations at
such wavelengths reveal the hot stellar content of these dust-enshrouded
environments like young HII regions in dense molecular clouds, the Galactic
centre or massive clusters. Recent examples which illuminate the advantages
of near- and mid-IR wavelengths for observing massive proto-stars and star
formation triggered by massive stars have been given by \citet{blum04},
\citet{whitney04} and \citet{clark04}.

Following the substantial progress in ground-based IR instrumentation
in the past decade, IR spectroscopy has become a powerful diagnostics for
the investigation of hot stars and the stellar winds surrounding them. The
first systematic {\it observational} studies of OB stars in the {\it H} and
{\it K} band have been performed by, e.g., \citet{hanson96}, \citet{Morris96}
and \citet{fn98} providing an important basis for quantitative spectral
analysis of early type stars. With the use of satellites (e.g., the Infrared
Space Observatory (ISO) in 1995 and the Spitzer Space Telescope in 2003) a
larger spectral window became accessible, completing the IR regime already 
observed from the ground.

{\it Modeling} of the near-infrared, on the other hand, has been performed 
mostly for early-type stars with {\it dense} winds, i.e., for Wolf-Rayet
Stars (\citealt{hil82}), Galactic centre objects (\citealt{naj94}), Of/WN
stars (\citealt{crow95, crow98}) and Luminous Blue Variables
(\citealt{naj97, naj98}). Combined ground-based optical and near-IR,
ISO and Spitzer mid-IR spectra for LBVs and Wolf-Rayets have been modeled by
\citet{najarro97b}, \citet{dessart00} and \citet{Morris00, Morris04}, in a
similar spirit as outlined below, namely to compare optical parameters 
with those obtained over the wider spectral range, partly including also the
UV. Note that all these investigations have been performed by means of the 
model atmosphere code {\sc cmfgen} (\citealt{hil98}).

For objects with thinner winds (which are of particular interest when aiming
at the youngest objects emerging from Ultra-Compact \HII\, (UCHII) regions),
no results are available so far, except from a pilot study by \citet{len04}.
In this study, various {\it synthetic} H/He IR-profiles, located in the {\it
J} to {\it L} band, are presented for a comprehensive grid of O-type stars
(from dwarfs to supergiants), and their diagnostic potential and value is
discussed. 

The reader may note that most of the available datasets of IR-spectra have
been observed at relatively low resolution (typically, at $R \approx
2,000$, though \citet{fn98} present a few spectra with $R \approx
10,000$), which compromises a precise spectroscopic analysis, since 
many decisive spectral features remain unresolved. Meanwhile, however,
\citet{hanson05} have re-observed a large sample of Galactic O-type
``standards'' with much higher resolution, typically at $R \approx 12,000$.
The objects where chosen in such a way that they both largely overlap with 
stars which have been analyzed before in the optical (e.g., \citealt{h02,
repo04}), and cover a wide range in spectral type and luminosity class.
Therefore, the present paper has the following objectives:

\begin{itemize}
\item[$\bullet$] 
We carry out a spectral analysis for this sample in the near infrared regime
and compare it with results already obtained in the optical. This will allow us
to check the extent to which the data derived from the IR is consistent with results
obtained from alternative studies in different wavelength bands. As an
ultimate goal, we plan to use solely the infrared regime to provide accurate
constraints to the characteristics of stars which can only be observed at
these wavelengths.

\item[$\bullet$]We test our model atmosphere code  {\sc fastwind} 
(\citealt{sph97, h02, puls05}) for OB stars in the near infrared, hence 
extending its usage to these wavelength ranges.

\item[$\bullet$] We give special attention to those lines 
which are located in the {\it H} and {\it K} band, i.e., which can be accessed by {\it
ground-based} instrumentation alone. Note that these lines are mainly 
formed close to the photosphere, i.e., remain uncontaminated by additional
effects such as clumping and X-rays and, thus, should provide rather robust
estimates for effective temperatures and gravities. 
\end{itemize}

The remainder of this paper is organized as follows. In Sect.~\ref{obs} we
briefly describe the observations and the lines used in our analysis. In
Sect.~\ref{modelcalc} we summarize our model calculations and comment on our
treatment of line-broadening for the hydrogen lines. Sect.~\ref{predictions}
outlines some theoretical predictions concerning the behaviour of strategic
lines, and Sect.~\ref{comp} compares our results with those obtained by
\citet{len04} by means of the alternative wind-code {\sc cmfgen} (see
above). In Sect.~\ref{analysis}, we discuss the analysis of the individual
objects of our sample, and in Sect.~\ref{logqapproach} we consider the 
consequences related to missing knowledge of stellar radius and 
terminal wind velocity. Sect.~\ref{compdata} compares
our IR results with those from the corresponding optical data. In
Sect.~\ref{summary}, finally, we present our summary and conclusions.

\section{Observations, targets and strategic lines}
\label{obs}

\begin{table}
\caption{Sample stars and observing data in the {\it H} and {\it K} band.
In sub-samples~I to III we have grouped those
objects which have been previously analyzed in the optical (sub-sample~I:
\citealt{repo04}; sub-sample~II: \citealt{h00, h02}; sub-sample~III: 
\citealt{kudetal99}). Subsample IV comprises those objects covered by 
various authors or not analyzed at all.}
\tabcolsep1.5mm
\renewcommand{\arraystretch}{1.1}
\begin{tabular}{l l c r}
\hline
Star & Sp.Type & {\sc subaru-ircs} & sample\\
\hline
Cyg\,OB2 $\#$7  & O3 If$^*$ & Nov 01      & II\\
Cyg\,OB2 $\#$8A & O5.5 I(f) & July 02     & II\\
Cyg\,OB2 $\#$8C & O5 If  & July 02        & II\\
HD\,5689   & O6 V      & Nov 01/July 02   & II \\
HD\,13268  & ON8 V     & Nov 01           & I\\
HD\,13854  & B1 Iab    & Nov 01           & III \\
HD\,13866  & B2 Ib     & July 02          & III\\
HD\,14134  & B3 Ia     & July 02          & III\\
HD\,14947  & O5 If+    & Nov 01           & I\\
HD\,15570  & O4 If+    & Nov 01           & II\\
HD\,15558$^{1)}$  & O5 III(f) & July 02   & I\\
HD\,15629  & O5 V((f)) & July 02          & I \\
HD\,30614  & O9.5 Ia     & Nov 01         & I\\
HD\,36166  & B2 V      & Nov 01           &  IV\\
HD\,37128  & B0 Ia     & Nov 01           & III\\
HD\,37468  & O9.5 V    & Nov 01           & IV\\
HD\,46150  & O5 V((f)) & Nov 01           & IV\\
HD\,46223  & O4 V((f)) & Nov 01           & IV\\
HD\,64568  & O3 V((f)) & Nov 01           & IV\\
HD\,66811  & O4 I(n)f  & Nov 01           & I\\
HD\,149438$^{1,2)}$ & B0.2 V    & July 02   & IV\\
HD\,149757 & O9 V    & July 02            & I\\
HD\,190864$^{2)}$ & O6.5 III(f)  & July 02       & I\\
HD\,191423 & O9 III:n$^*$ & July 02       & I\\
HD\,192639 & O7 Ib     & July 02          & I \\
HD\,203064 & O7.5 III:n ((f)) & July 02   & I\\
HD\,209975 & O9.5 Ib   & July 02          & I\\
HD\,210809 & O9 Iab    & July 02          & I \\
HD\,217086 & O7 Vn     & Nov 01           & I \\
\hline
\end{tabular}
\smallskip
\newline
$^{1)}$ only {\it K} band available.\\
$^{2)}$ Additional {\sc irtf-cshell} spectra covering \HeI2.05 available.\\
\label{runs}
\end{table}

For our analysis we use a subset of stars given by \citet{hanson05}.
Detailed information about the observation dates, resolution, spectrometers and
data reduction can be found there. We selected the spectra from that
sample which where obtained with the Infrared Camera and Spectrograph 
({\sc ircs}) mounted at the Cassegrain focus of the 8.2m Subaru Telescope 
at Mauna Kea, Hawaii. This totaled in 29 stars out of the 37 targets
collected by \citet{hanson05}. 

The targets had been selected i) to adequately cover the complete OB star range
down to B2/B3 at all luminosity classes, and ii) that most of them have already
been analyzed in the optical (for details, see \citealt{hanson05}).
According to the purpose of our analysis, we have
exclusively used the data from the Subaru Telescope and not the VLT data
(comprising the remaining 8 objects), since we did not possess complementary
optical spectra for the latter dataset. In the following, we will define
four different sub-samples denoted by I to IV in order to distinguish
between objects analyzed in the optical by different authors. Sample~I
comprises those stars discussed by \citet{repo04}, sample~II corresponds to
objects analyzed by \citet{h00, h02}\footnote{Note that the first of the two
investigations has been performed by unblanketed models.}, sample~III
(B-supergiants) has been analyzed by \citet[ only with respect to
wind-parameters]{kudetal99}, and sample~IV consists of the few remaining
objects considered by various authors or not at all. In particular,
HD\,46150 has been investigated by \citet[ plane-parallel, unblanketed
models]{h92} and $\tau$~Sco (HD\,149438) by \citet[ plane-parallel NLTE
analysis with underlying Kurucz models]{Kilian91} and by \citet{PB04} with
respect to optical and IR hydrogen lines. Table \ref{runs} indicates to
which individual sub-sample the various objects belong.

The Subaru/{\sc ircs} {\it H} band and {\it K} band spectral resolution is R
$\approx$ 12000. The typical signal-to-noise ratios obtained with these
spectra were S/N $\approx $ 200-300, with areas as high as S/N $\approx $
500, and as low as S/N $\approx $ 100, depending on the telluric
contamination. The spectra were obtained over two separate runs, the first
in November 2001 and the second in July 2002. Due to poor weather condition,
the telluric corrections for some of the spectra proved to be difficult.
This can be seen in the {\it H} band spectra of HD\,217086, HD\,149757,
HD\,66811, HD\,5689 and HD\,15629. Furthermore, there were no {\it H} band
spectra of HD\,15558 and $\tau$ Sco available, weakening the significance of
their analyses. The reduction of the data was performed using {\sc iraf}
routines and Perl {\sc idl} including standard procedures such as bias
subtraction, flat field division, spectrum extraction, wavelength
calibration and continuum rectification. Table~\ref{runs} summarizes all
observational runs obtained with {\sc ircs}. In the following, all wavelengths
of NIR lines are given in microns ($\mu$m).

The data for the \HeI \lam 2.05 line, which had not been observed by {\sc
subaru} were taken at the Infrared Telescope Facility ({\sc irtf}) in March,
June and July of 2003.  The {\sc cshell} echelle spectrograph
(\citealt{green93}) was used with a slit of 1.0 arcseconds. The instrumental
spectral resolving power as measured by a Gaussian fit to the OH night sky
emission lines was 4.0 pixels FWHM, or 12.1 \kms, corresponding to a
resolution of 24,000.  The spectra were reduced using {\sc iraf} routines
and the subsequent analysis was done using routines written in Perl {\sc
idl}. For all spectra, dark frames and flat field frames were averaged
together to form a master dark and flat frame. Unfortunately, \HeI2.05 lies
within a region where the telluric absorption is extremely large, degrading
the signal significantly (\citealt{kenw04}). After the reduction, it turned out that most of
our spectra did not posses sufficient quality (only moderate S/N), and we
could use only the spectra obtained for two of the stars (HD\,190864 and
$\tau$ Sco) for our analysis.  Nevertheless, in all cases we have included the
{\it synthesized} line for the sake of completeness. 

The spectral classification of sample I is the one adopted by \citet{h92},
based mostly on the work by \citet{wal72, wal73}, the unpublished catalogue
of OB stars by C.~Garmany and by \citet{math89}. As for samples II to IV,
the spectral classification used by \citet{hanson05} has been retained.  The
classifications were based mostly on Walborn classifications, except for the
Cyg OB2 stars, which relied on \citet{masthom91}.

\medskip 
\noindent 
In total the sample consists of 29 Galactic O and early B type stars as
listed in Table~\ref{runs} ranging from O3 to B3 and covering luminosity
class Ia/Iab, Ib/II, III, and V objects, where 4 stars (of the latest
spectral types) have been discarded later in the study. The strategic lines
used in our analysis are (all wavelengths in air)
\begin{itemize}
\item[$\bullet$]{\it H}~band 
\begin{itemize}
\item \HI\ \lam 1.68 (n = 4 \rarrow 11, Br11), 
\item \HI\ \lam 1.74 (n = 4 \rarrow 10, Br10),
\item \HeI\ \lam 1.70 ({$3p$ $^{3}{\rm P}^{\rm o}$} - {$4d$ $^{3}{\rm D}$}, 
triplet), 
\item \HeII\ \lam 1.69 (n = 7 \rarrow 12). 
\end{itemize}
\item[$\bullet$]{\it K}~band
\begin{itemize}
\item \HI\ \lam 2.166 (n = 4 \rarrow 7, \Brg),
\item \HeI\ \lam 2.058 ({$2s$ $^{1}{\rm S}$} - {$2p$ $^{1}{\rm P}^{\rm o}$},
singlet), where available,
\item \HeI\ \lam 2.11 (comprising the \HeI\ triplet  
\lam 2.1120 ({$3p$ $^{3}{\rm P}^{\rm o}$} - {$4s$ $^{3}{\rm S}$}) 
and the \HeI\ singlet 
\lam 2.1132 ({$3p$ $^{1}{\rm P}^{\rm o}$} - {$4s$ $^{1}{\rm S}$})), 
\item \HeII\ \lam 2.188 (n = 7 \rarrow 10). 
\end{itemize}
\end{itemize}
Note that \Brg\ overlaps with the \HeI\ triplet 
\lam 2.1607 ({$4d$ $^{3}{\rm D}$} - {$7f$ $^{3}{\rm F}^{\rm o}$}), 
the \HeI\ singlet 
\lam 2.1617 ({$4d$ $^{1}{\rm D}$} - {$7f$ $^{1}{\rm F}^{\rm o}$}) 
and \HeII\ \lam 2.1647 (n = 8 \rarrow 14). Whereas the singlet is 
not included in our formal solution, the \HeI\ triplet, in particular, 
has been used to check the consistency of our results. Note that the
influence of the \HeII\ lines overlapping with Br10 and Br11 is marginal.

\section{Model calculations}
\label{modelcalc}

The calculations presented in this paper have been performed by means of our
present version of {\sc fastwind}, as described by \citet{puls05}. In
addition to the features summarized in \citet{repo04}, this code meanwhile
allows for the calculation of a consistent\footnote{Note, however, that
non-radiative heating processes might be of importance.} temperature,
utilizing a flux-correction method in the lower atmosphere and the thermal
balance of electrons in the outer one. As has been discussed, e.g., by
\citet{kubat99}, the latter method is advantageous compared to exploiting
the condition of radiative equilibrium in those regions where the radiation
field becomes almost independent on $\Te$. Particularly for
IR-spectroscopy, such a consistent T-stratification is of importance, since
the IR is formed above the stellar photosphere in most cases and depends
(sometimes critically) on the run of $\Te$. We have convinced ourselves that
our previous results concerning optical lines remain (almost) unaffected by
this modification. 

\citet{puls05} present a thorough comparison with models from alternative
``wind-codes'' ({\sc wm}-basic, \citealt{Paul01} and {\sc cmf\-gen},
\citealt{hil98}). Some differences were seen in the O{\sc ii} continuum at
and below 350 \AA\ ({\sc fastwind} predicts a higher degree of line-blocking
in this region), which might have some influence on the helium ionization
balance, due to a different illumination of the \HeII\ resonance lines.
Also, {\sc cmfgen} predicted weaker optical \HeI\ singlets in the
temperature range between 36,000 to 41,000~K for dwarfs and between 31,000
to 35,000~K for supergiants. Otherwise, the comparison resulted in very good
agreement.

\subsection{Atomic data and line broadening}
\label{atomdat}

In order to obtain reliable results in the IR, our present H and \HeII\
models consist of 20 levels each, and \HeI\ includes levels until $n=10$,
where levels with $n=8, 9, 10$ have been packed.  Further information
concerning cross-sections etc. can be found in \citet{jok02}. 

The hydrogen bound-bound collision strengths require some special remarks.
The atomic data on {\it radiative} line processes in \HI\ are very
accurate because they can be obtained analytically due to the two-body
nature of the hydrogen atom. However, for excitation/de-excitation processes,
these involve a colliding particle, making the situation much more complex.
In most cases only approximation formulae are available. 

Note that the ``choice'' of the collisional data is an especially important
factor for the line formation in the IR. Although the effect of different
collisional data will not be apparent for the ground state, higher levels
display a significant sensitivity, reaching its maximum for levels with
intermediate {\it n} at line formation depth. Recently, \citet{PB04} have 
emphasized the differences in the collisional cross section from
approximation formulae and ab initio computations for transitions up to $n =
7$. Particularly, the frequently used approximations by \citet{mha75} and by
\citet{j72} show a different behaviour and fail to simultaneously reproduce
the optical and IR spectra over a wide parameter range.  However, the
collisional data provided by Przybilla \& Butler (in combination with the
approximation formulae by \citet{PR78} and \citet{mha75}) are able to
reproduce the observed line profiles in those cases which have been checked.
Note, however, that these checks did not cover O-type supergiants,
cf. Sect.~\ref{collhyd}!

The standard implementation of the corresponding cross sections in {\sc
fastwind}, on the other hand, is based on data presented by \citet{gio87}.
Although affected by similar problems as described above, the differences to
the ab initio calculations are smaller but still worrisome. As detailed
later on, a comparison of simulations using both data-sets alternatively
revealed that {\it for our O-star sample} we find better agreement with
corresponding optical results if our standard implementation is used.
Consequently, all calculations described in the following are based on these
data, whereas further comments concerning the effect of incorporating the
data by \citet{PB04} are given in Sect.~\ref{collhyd}.

Since we are concentrating on those lines which are formed close to the
photosphere, line-broadening is particularly important (and leads to a number
of interesting effects, shown below). Unfortunately, calculations as ``exact'' 
as for optical lines do not yet exist for their IR counterparts, leaving us 
to use reasonable approximations. 

Actually, \citet{lemke97} has published extended Stark broadening tables
(based on the approach by \citealt{vcs73}, ``VCS'') for the hydrogen Lyman
to Brackett series. In a first step, we have used his dataset for the
calculation of the Brackett lines.  However, we immediately realized that at
least Br11 must be erroneous by virtue of a comparison with observed
mid-resolution B-type NIR spectra which revealed no problems if approximate
broadening functions are used (see \citealt{hansonetal03}, Fig.~4).  After a
careful investigation by K.~Butler (priv. comm.), it turned out that not only
Br11 but also other transitions, i.e., predominantly members of the higher
series, are affected by a number of (numerical) problems in the code used by
Lemke.

Thus, Stark-broadening of hydrogen needs to be approximated as well.  We
follow the method by \citet{griem67} as outlined in \citet[Appendix]{am72},
based on a corrected asymptotic Holtsmark formula. Due to comparisons with
VCS calculations for optical transitions from \citet{schoening89}, which are
used by {\sc fastwind} anyway, we have convinced ourselves that the Griem
approximation recovers the more exact VCS results with very high precision,
if the upper and lower level of the transition lie not too closely together
(e.g., H$_\alpha$ is badly approximated, whereas for H$_\gamma$ no
differences are visible). The results obtained by using either the
(erroneous) data by Lemke or the Griem approximation are given by means of a
detailed comparison later on, cf. Fig.~\ref{lemkegriem}. Griem-broadening is
also applied to \HeII\ (1.69, 2.18 $\mu$m), whereas for \HeI\ (1.70, 2.05, 2.11
$\mu$m) we have used Voigt functions only, with damping parameters accounting
for natural and collisional broadening. The comparison to observations
suggests that this approximation describes reality sufficiently well.

\section{Predicted behaviour of strategic lines}
\label{predictions}

\begin{figure}
\resizebox{\hsize}{!}
   {\includegraphics{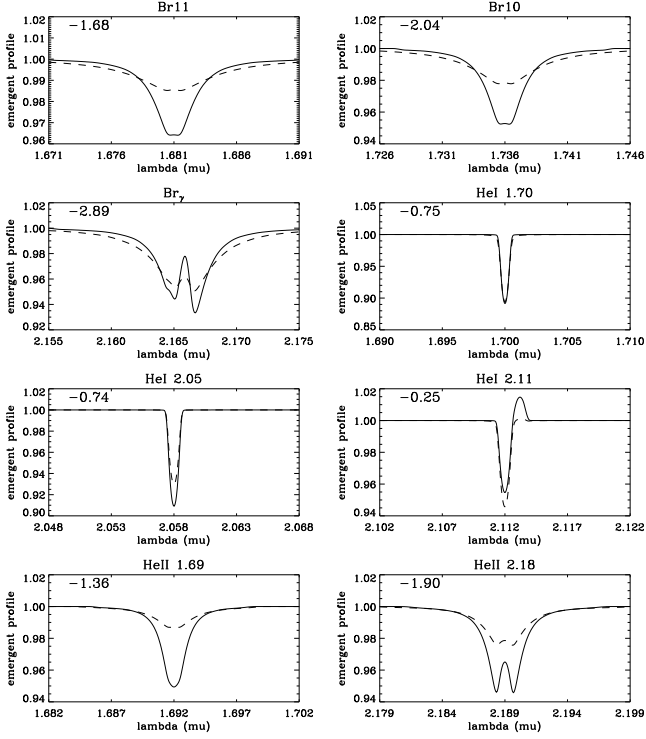}}
\caption{Comparison of strategic NIR lines for two atmospheric models at
\teff = 40,000~K and different gravities, \g=3.7 (solid) and \g=4.5
(dashed), respectively. Both models have a negligible wind, with $\log Q =
-14$. The number in the upper left corner gives the equivalent width (in
\AA) of the low-gravity model, where, in agreement with previous papers,
negative numbers indicate net-absorption. All profiles are displayed on the
same horizontal scale (of width 0.02$\mu$m), and the profiles have been
rotationally convolved with \Vr = 80 \kms.}
\label{compgrav}
\end{figure}

Before we describe the results of our analysis, we will investigate the
behaviour of our synthesized lines in some detail, particularly because their
dependence on gravity seems to be somewhat strange, at least if one
extrapolates the knowledge accumulated in the optical. Although a related
investigation has already been performed by \cite{len04}, they have only 
discussed the behaviour of the equivalent widths. Moreover, their model
grid is rather restricted and does not allow the investigation of changes in
synthesized profiles if only {\it one} atmospheric parameter is altered. On
the other hand, we have calculated a rather large grid of models in the
parameter range $20,000 < \Teffe < 50,000$ with a typical variation in $\log
g$ over two dex, and wind strengths varying from negligible to very large
(cf. \citealt{puls05}), allowing us to inspect this kind of reaction in more
detail. In the next section we will, of course, compare our results
also to those obtained by \citet{len04}.

As a prototypical example, in Fig~\ref{compgrav} we compare the strategic
H/He NIR lines for a model at \teff = 40000~K, with \g = 3.7 (solid) and 
4.5 (dashed). Both models have a vanishing wind density, corresponding to
$\log Q = -14$, where $Q$ is a suitable measure to compare the influence of
different wind strengths in recombination lines (see \citet{puls96}).
Throughout this paper, we have defined

\beq Q=\frac{\Mdote[\Msune/{\rm yr}]}{\bigl((\Rstare/\Rsune) \vinfe [\rm{kms^{-1}}]\bigr)^{1.5}}.
\label{defQ} \eeq 

Most interestingly, almost all NIR features {\it become deeper and their
equivalent width increases if the gravity decreases}. In contrast to the 
Balmer lines, the cores of Br10, Br11 (and of \HeII1.69/2.18) are 
significantly anti-correlated with gravity. This behaviour is
completely opposite to what one expects from the optical. Only the far wings
of the hydrogen lines bear resemblance to the optical, which become shallower
when the gravity decreases.

\begin{figure*}
\begin{minipage}{8.8cm}
\resizebox{\hsize}{!}
   {\includegraphics{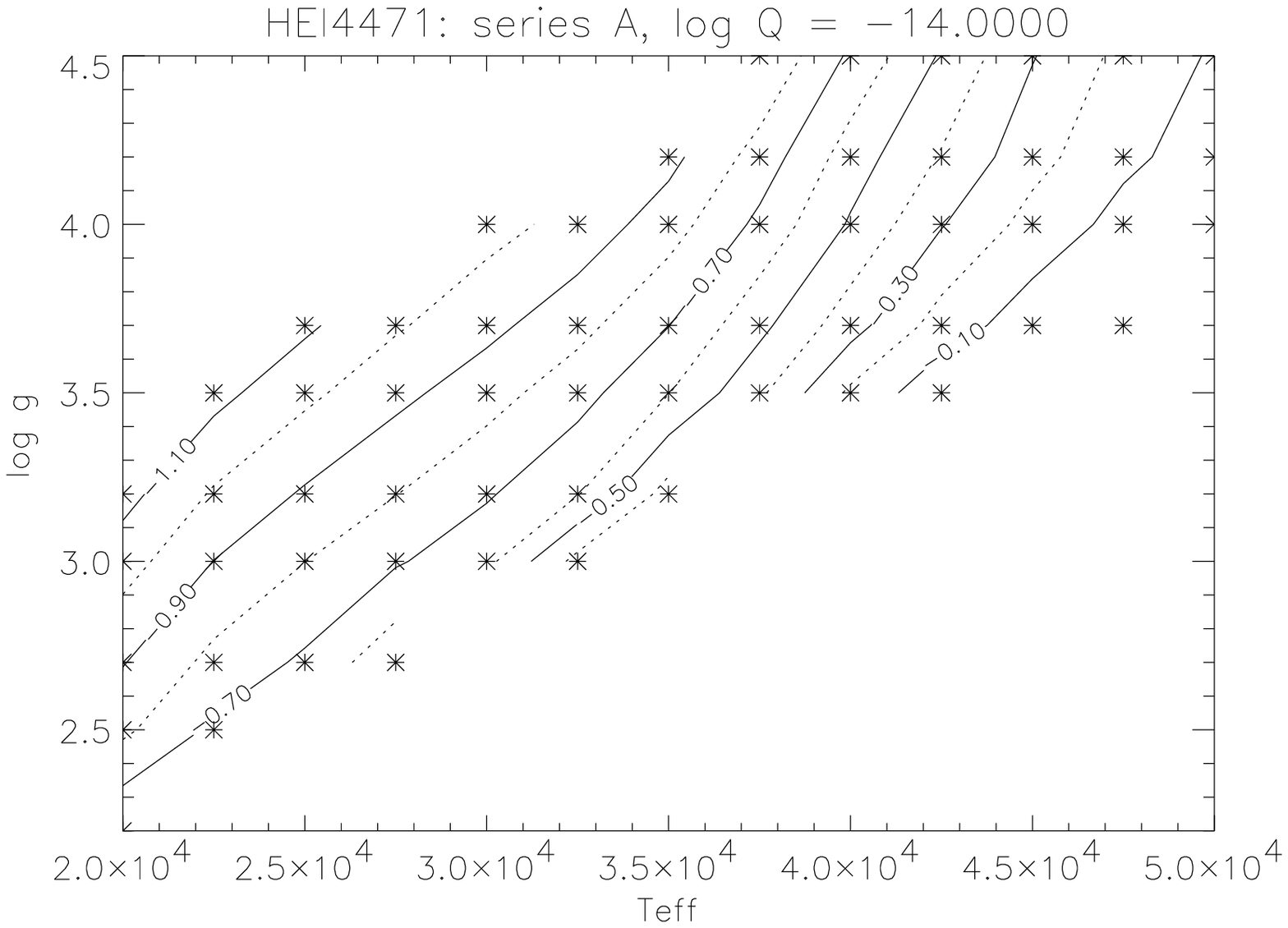}}
\end{minipage}
\hfill
\begin{minipage}{8.8cm}
   \resizebox{\hsize}{!}
      {\includegraphics{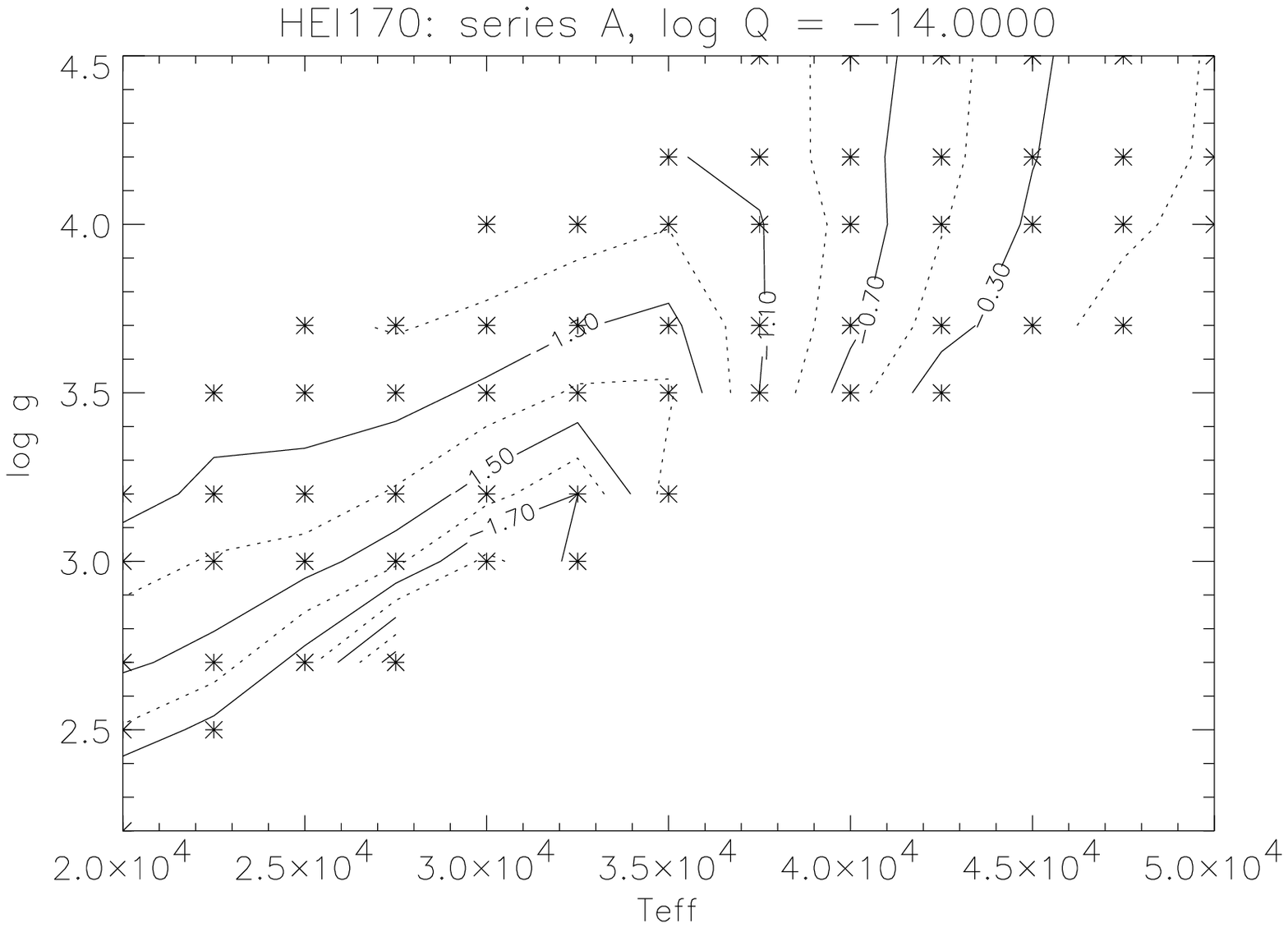}}
\end{minipage}
\begin{minipage}{8.8cm}
\resizebox{\hsize}{!}
   {\includegraphics{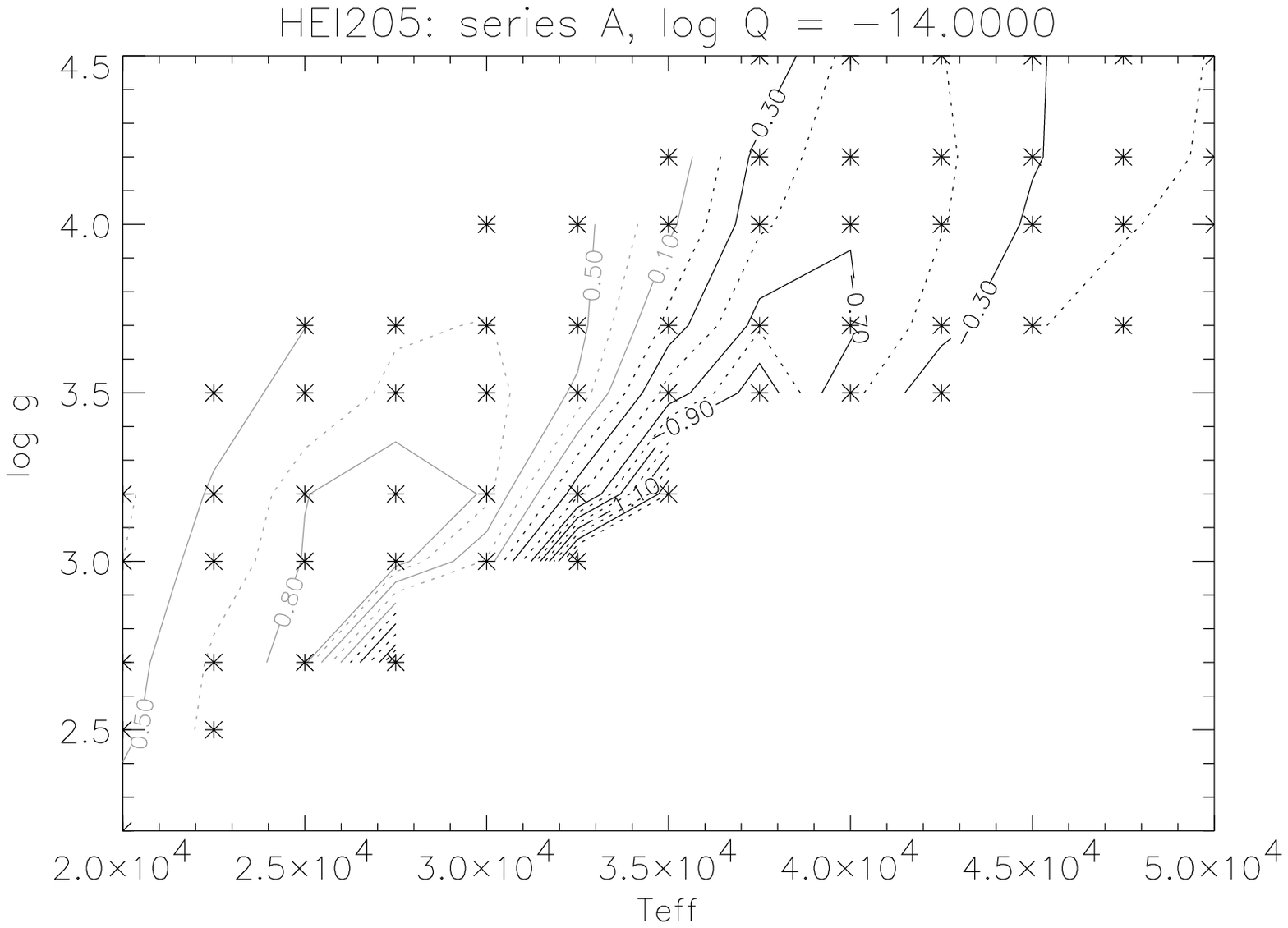}}
\end{minipage}
\hfill
\begin{minipage}{8.8cm}
   \resizebox{\hsize}{!}
      {\includegraphics{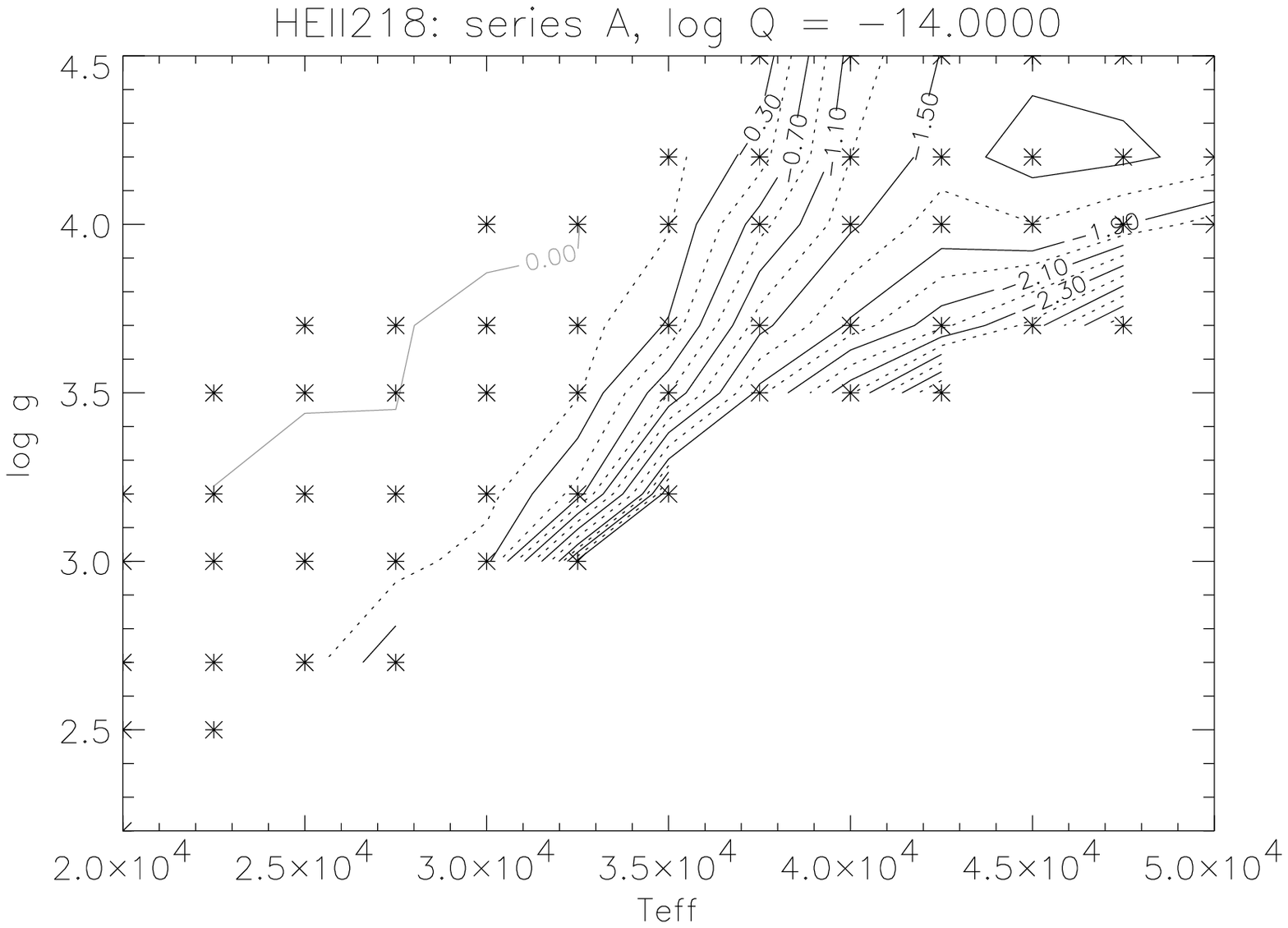}}
\end{minipage}
\caption{Iso-contours of equivalent widths for \HeI\ and \HeII\ lines using a
model-grid with values for \teff\ and \g\ as indicated, and negligible
wind densities, $\log Q = -14$. Curves in grey color (left-hand side of
\HeI2.05 iso-contours) indicate net emission. Note that for the optical
transition (\HeI4471) the absorption increases as function of gravity,
whereas for the NIR lines this behaviour is mostly reversed.
Asterisks denote the position of the calculated models (see also
\citealt{puls05}).} 
\label{isocontours}
\end{figure*}

Although the reaction of \HeI\ on \g\ is only moderate, at lower 
temperatures (with more \HeI\ present) we observe the same trend, i.e., the
equivalent width (e.w.) increases with decreasing gravity, as shown in the
e.w. iso-contour plots in Fig.~\ref{isocontours}. For comparison, this plot
also shows the extremely ``well-behaved'' \HeI4471 line, which decreases in
strength with decreasing gravity in all regions of the \teff-\g\ -plane. 

Before we will further discuss the origin of this peculiar behaviour of
NIR-lines, let us point out that these trends do {\it not} depend on
specific details of the atmospheric model, particularly not on the presence
or absence of a temperature inversion in the upper photospheric layers. The
same relations (not quantitatively, but qualitatively) were also found in
models with a monotonically decreasing temperature structure in the inner
part ($\log \taur > -2$) and a constant minimum temperature in the outer
wind.

\begin{figure*}
\begin{minipage}{8.8cm}
\resizebox{\hsize}{!}
   {\includegraphics{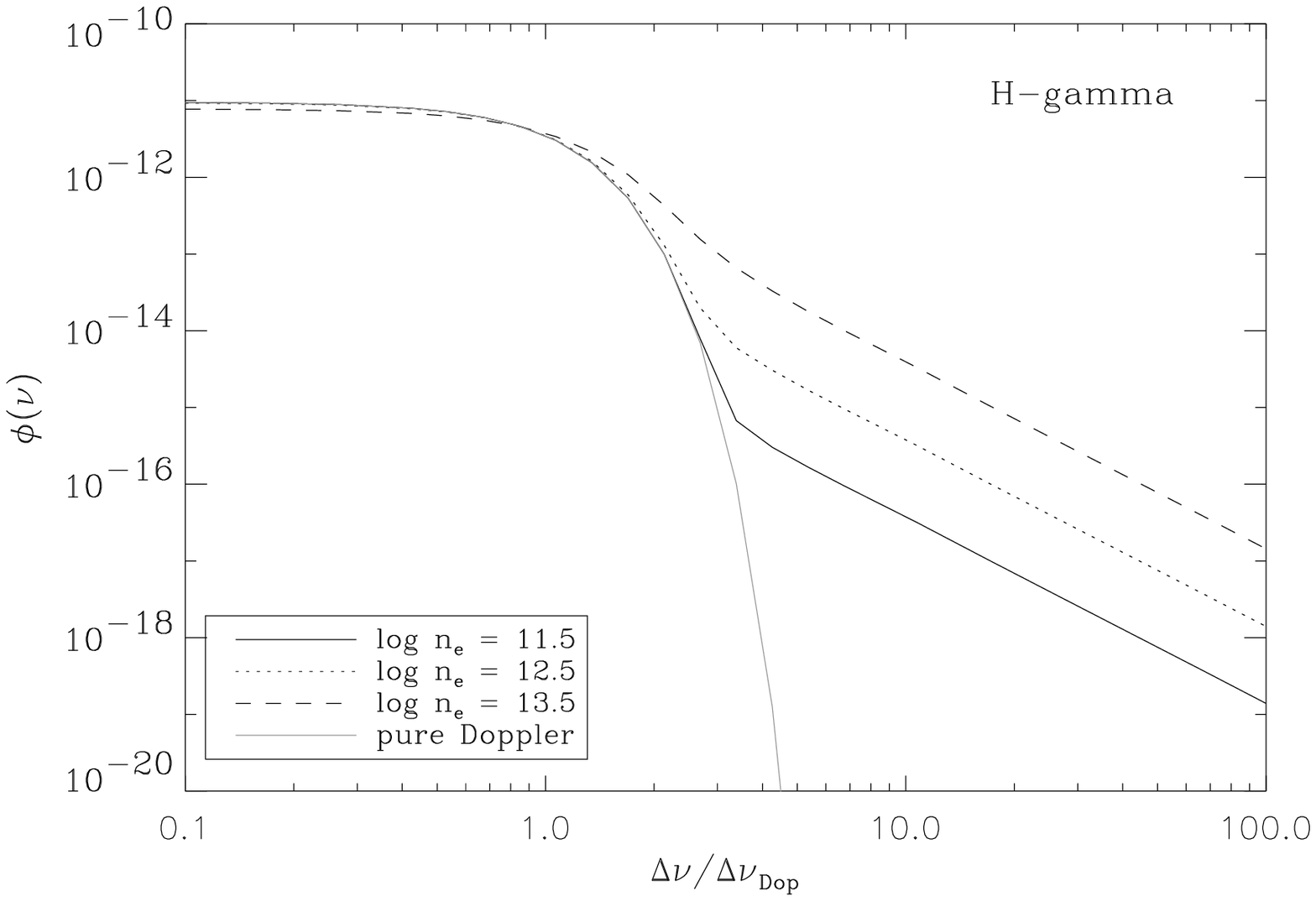}}
\end{minipage}
\hfill
\begin{minipage}{8.8cm}
   \resizebox{\hsize}{!}
      {\includegraphics{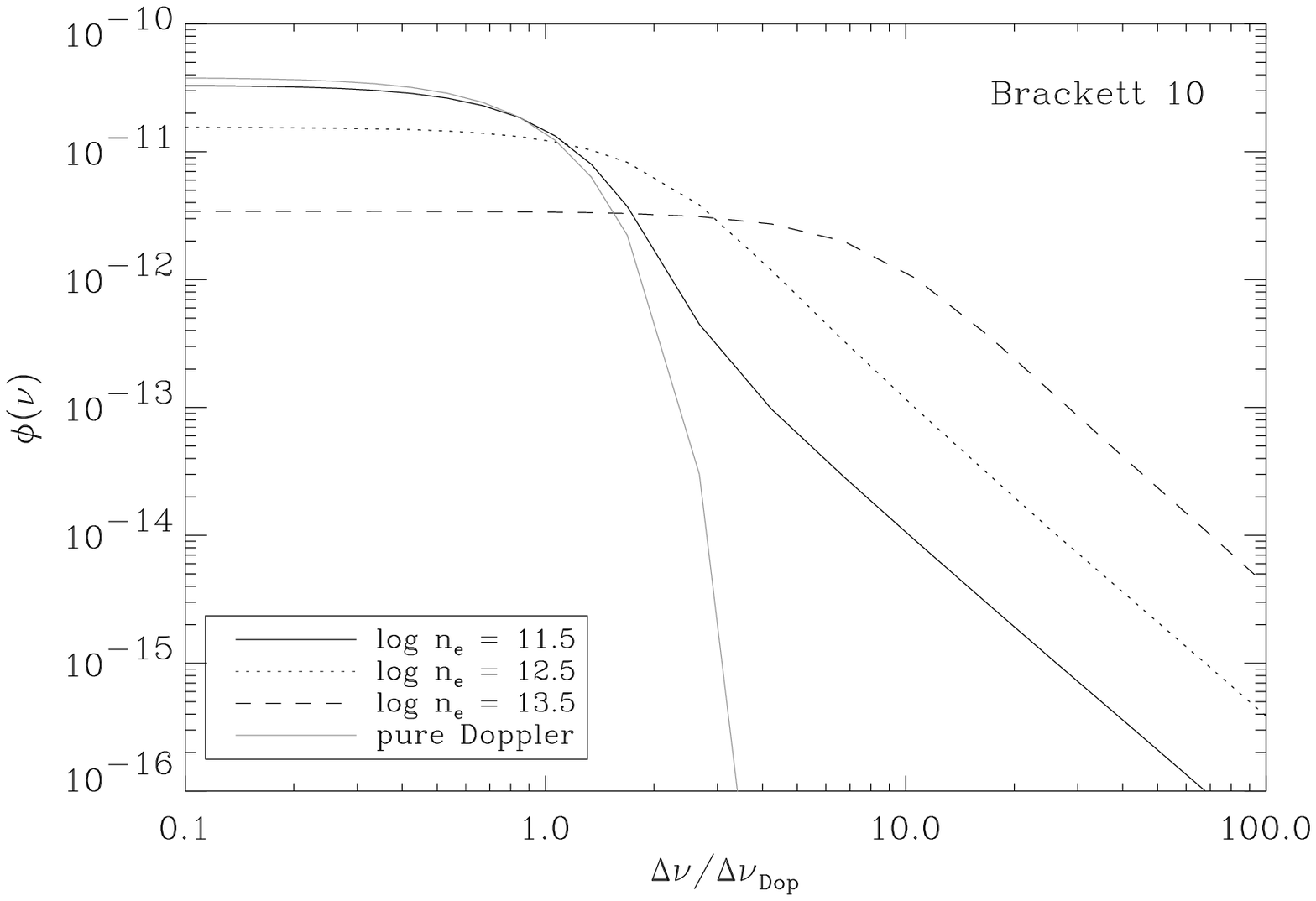}}
\end{minipage}
\caption{Line broadening profile functions (convolution of Doppler with
Stark profiles, in [s$^{-1}$]) for \Hg\ and Br10, as a function of frequency
displacement from the line centre in units of thermal Doppler-width. Both
profiles have been calculated in Griem approximation, for an electron
temperature $\Te$ = 40,000~K and three different electron densities typical
for the line forming region. The grey line corresponds to a pure Doppler
profile. Note that for \Hg\ the core of the Stark profile is identical with
the pure Doppler profile, whereas for Br10 the Stark core is extremely
density dependent and coincides with the Doppler profile only at lowest
densities (see text).} 
\label{starkprofiles}
\end{figure*}

\subsection{Hydrogen and \HeII\ lines: Influence of Stark broadening}
\label{stark}

The peculiar behaviour of the line cores of the hydrogen Brackett lines and
\HeII1.69/2.18 can be understood from the reaction of the core of the
corresponding Stark-profiles as a function of electron density.
Fig.~\ref{starkprofiles} shows the Stark-profiles for \Hg\ and Br10 as a
function of frequency displacement from the line centre in units of thermal
Doppler-width, calculated in the Griem approximation. Both profiles have been
calculated for typical line-forming parameters, $\Te$ = 40,000~K and $\log
n_{\rm e} = 11.5, 12.5$ and 13.5, respectively. The corresponding pure
Doppler profile is overplotted (in grey). The decisive point is, that for
\Hg, with relatively low upper principal quantum number, the Stark width is
not considerably large, and the core of the profile is dominated by
Doppler-broadening, independent of electron density. Only in the far wings
does the well known dependence on $n_{\rm e}$ become visible. On the other hand,
for Br10 the Stark width becomes substantial (being proportional to the
fourth power of upper principal quantum number), and even the Stark-core
becomes extremely density dependent. Only at lowest densities, the profile
coincides with the pure Doppler profile, whereas for larger densities the
profile function (and thus the frequential line opacity) decreases with
increasing density. In the far cores, finally, the conventional result
($\phi(\nu)$ correlated with $n_{\rm e}$) is recovered. Thus, as a
consequence of the dependence of Stark-broadening on density, the line cores
of the hydrogen lines with large upper principal quantum number become
weaker with increasing gravity. Br$_\gamma$
(with upper quantum number n=7) is less sensitive to this effect, cf.
Fig~\ref{compgrav}. 

\begin{figure}
\resizebox{\hsize}{!}
   {\includegraphics{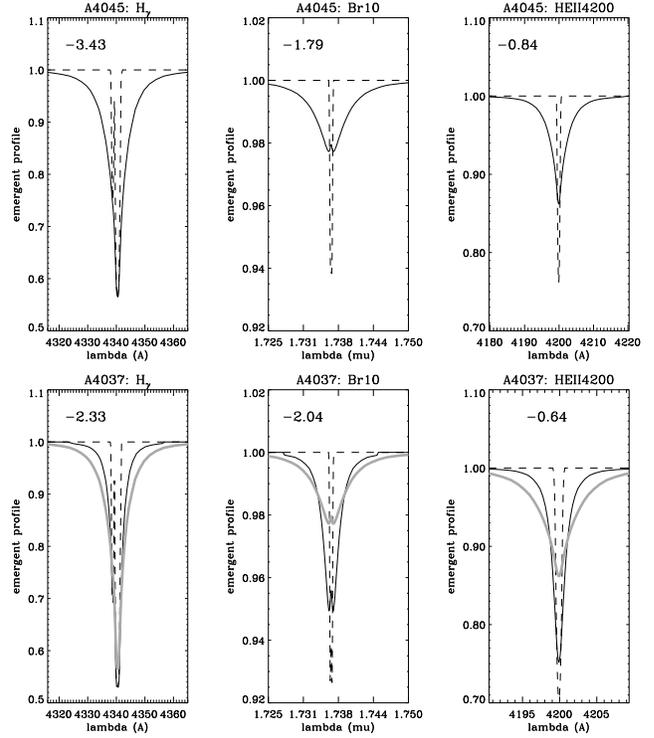}}
\caption{Influence of Stark-broadening for lines with low- and high-lying 
upper level as a function of gravity. Upper level, solid lines: 
Synthetic spectra of
\Hg, Br10 and \HeII4200 (n=4 \rarrow 11) for an atmospheric model with \teff =
40,000~K, \g=4.5 and $\log Q = -14$ (cf. dashed lines in Fig.~\ref{compgrav}).
Overplotted (dashed) are the corresponding profiles with pure
Doppler-broadening. Lower panel: As the upper panel, but for an atmosphere
with \g=3.7 (solid lines in Fig.~\ref{compgrav}). For comparison, the results for
the Stark-broadened profiles from the upper panel are overplotted in grey.
Note the similarity in effects between Br10 and \HeII4200.}
\label{compdoppler}
\end{figure}

In Fig.~\ref{compdoppler}, we demonstrate the different reactions of the
Stark profiles on electron density (gravity) by comparing the synthesized
{\it emergent} profiles of the high- and low-gravity model at
\teff=40,000~K, as described above. In particular, we compare these profiles
with the corresponding profiles calculated with pure Doppler-broadening. For
\Hg, the core of the Stark-broadened profile agrees well with the
Doppler-broadened one (dashed) in both cases. The major difference is
found in the far wings, which become wider and deeper as a function of
(electron-) density, thus, being the most useful indicators for the effective
gravity. For Br10, on the other hand, the pure Doppler profile is much
deeper than the Stark-broadened core, where the differences are more
pronounced for the high gravity model. Note particularly that the (absolute)
e.w. is larger for the low gravity model (although the high gravity model
has more extended wings), since the major part of the profile is dominated
by the core which is deeper for lower gravities!

Actually, the same effect is already visible in the optical, namely for the
prominent \HeII\ lines at 4200\AA\ (transition 4-11) and 4541\AA\
(transition 4-9, not shown here). The increase in absolute e.w. as a
function of gravity is solely due to the wings. In accordance with
Br10/Br11, however, the cores of the lines become shallower with increasing
gravity, {\it not because of an effect of less absorbers, but because of
less frequential opacity due to a strongly decreased broadening function.}

Let us allude to an interesting by-product of our investigation. A
comparison of our synthetic NIR profiles with the observations will show
that in a number of cases the observed Br10/Br11 profiles cannot be fitted
in parallel. In this case the line formation is well understood and the
profiles from {\sc cmfgen} are identical (note that also the optical
hydrogen lines agree well, see \citealt{repo04}), giving us confidence that
our occupation numbers are reasonable and that the obvious differences are
due to inadequate broadening functions. 

\begin{figure*}
\begin{minipage}{8.8cm}
\resizebox{\hsize}{!}
   {\includegraphics{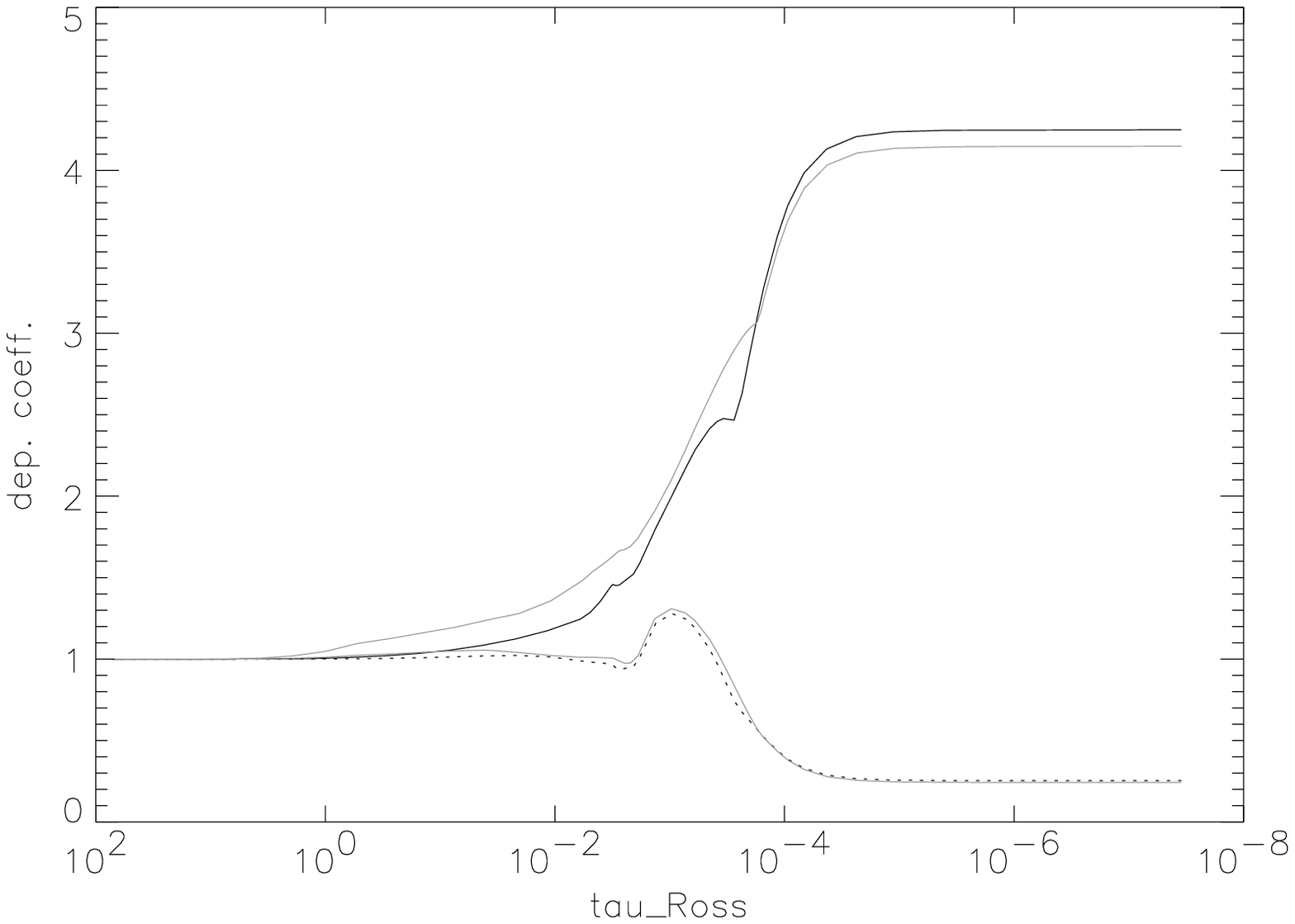}}
\end{minipage}
\hfill
\begin{minipage}{8.8cm}
   \resizebox{\hsize}{!}
      {\includegraphics{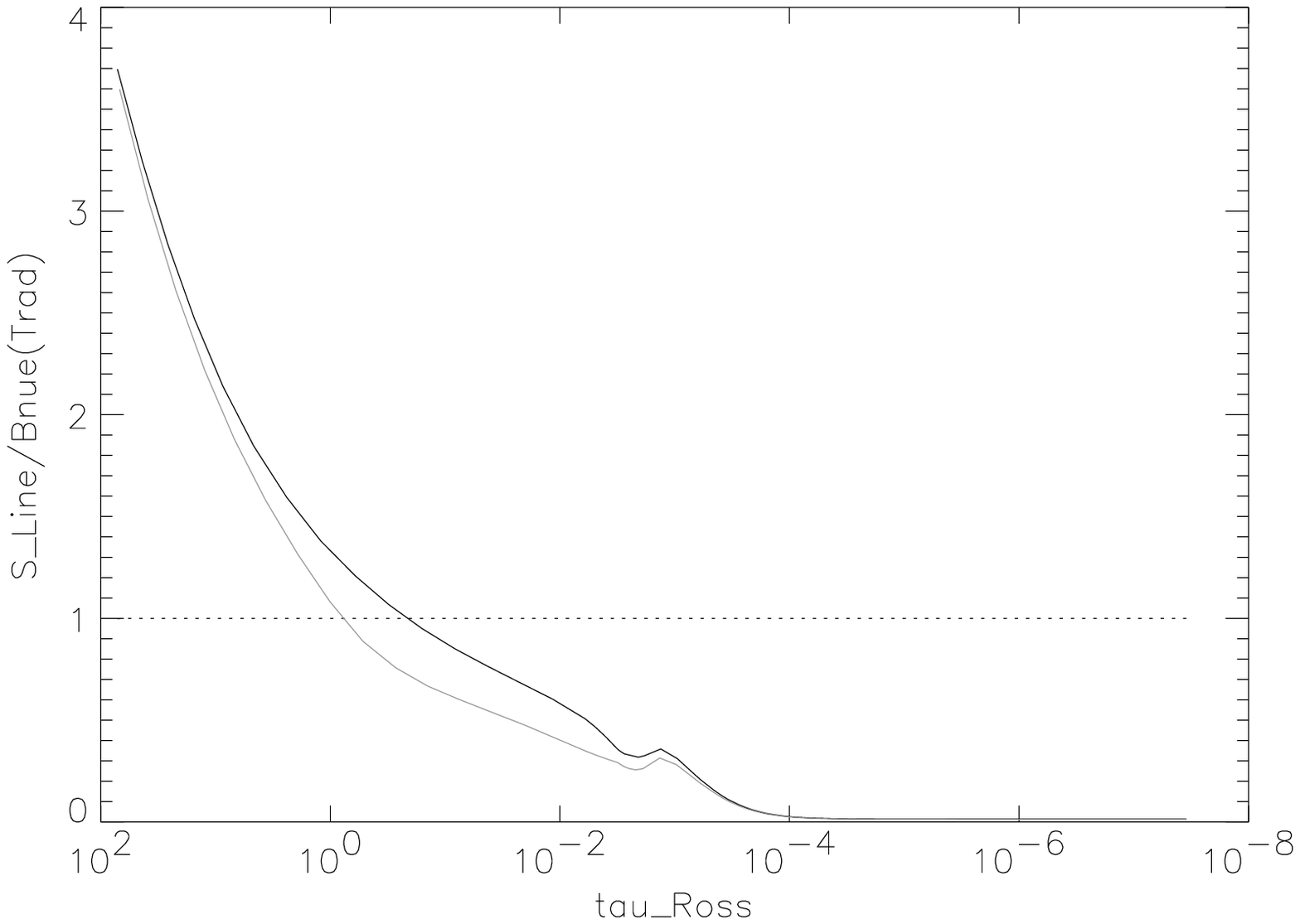}}
\end{minipage}
\caption{Left panel: NLTE-departure coefficients for the lower (solid) and
upper (dotted) levels of \HeI1.70 for a model with \teff = 30,000~K, \g=3.4
and negligible wind. Overplotted in grey are the corresponding values for a
similar model, but with lower gravity, \g=3.0. \newline Right panel: As
the left panel, but for the corresponding line source functions 
in units of the emergent continuum at 1.70$\mu$m.}
\label{hei170}
\end{figure*}

On the other hand, since \HeII4200/4541 is affected by
almost identical line broadening, we would like to suggest a solution for a
long standing problem in the {\it optical} spectroscopy of hot stars: It is
well known that for a wide range of O-star parameters the theoretical
simulations of these lines (by means of both plane-parallel and extended
atmospheres) have never been able to reproduce the observations in parallel
(e.g., \citealt{h02}), where the largest discrepancies have been found in
the line cores. The origin of this discrepancy is still 
unknown.\footnote{Note that this problem is most probably not related to the
presence of the N{\sc iii} blend in \HeII4200, since it occurs also in hot
objects.} Due to the similarity of this problem to the one shown by
Br10/11 and accounting for the similar physics, we suggest that also in this
case we suffer from an insufficient description of presently available
broadening functions (which are described within the
VCS-approach, see \citealt{schoening89}). Thus, a re-investigation of
line broadening for transitions with high lying upper levels seems to be
urgently required.

In summary, due to their tight coupling with electron density, the cores of
Br10/11 and \HeII1.69/2.18 are excellent indicators of
gravity, where deeper cores indicate lower gravities (if the (projected)
rotational velocities are similar). 

\subsection{\HeI\ lines: Influence of NLTE effects}
\label{nlteeffects}

The peculiar behaviour of the hydrogenic lines could be traced down to the
influence of the profile-functions, whereas the formation of most of the NIR
\HeI\ lines is dominated by NLTE-effects. As has been extensively discussed
by \citet{mih78}, \citet{kud79}, \citet{naj98}, \citet{PB04} and 
\citet{len04}, the low value of $h\nu/kT$ leads to the fact that even small
departures from LTE become substantially amplified in the IR (in contrast to
the situation in the UV and optical). A typical example is given by the
behaviour of \HeI1.70 at temperatures below \teff $\approx$ 35,000~K, cf.
Fig.~\ref{isocontours} (note, that {\sc cmfgen} gives identical
predictions). Again, this line becomes stronger for lower gravity, in
contrast to the well known behaviour of optical lines (compare with the
\HeI4471 iso-contours).

Fig.~\ref{hei170} gives a first explanation, by means of two atmospheric
models with \teff=30,000~K, \g=3.4 and 3.0, respectively, and (almost) no
wind. The departure coefficient of the upper level, $b_u$, of this
transition ($4d$ $^{3}{\rm D}$) is independent of gravity and has, in the
line forming region, a value of roughly unity (strong coupling to the \HeII\
ground-state).  The lower level of this transition ($3p$ $^{3}{\rm P}^{\rm o}$) is quite
sensitive to the different densities, i.e., being stronger overpopulated in the
low-gravity model. Consequently, the line source function, being roughly
proportional to $b_u/b_l$, is considerably lower throughout the photosphere
(right panel of Fig.~\ref{hei170}), and thus the profile is deeper, even if
the formation depth is reached at larger values of $\taur$.

The reason for this stronger overpopulation at lower \g-values is explained
by considering the most important processes which populate the $3p$-level.
First, the influence of collisions is larger at higher densities, which
drives the departure coefficient into LTE. Second, the level is strongly
coupled to the triplet ``ground state'' (i.e., the lowest meta-stable state)
which, in the photosphere, is overpopulated as an inverse function of the
predominant density. The overpopulation (with the consequence of
over-populating the $3p$-level) is triggered by the strength of the
corresponding ionizing fluxes. These are located in the near UV (roughly at
2,600 \AA) and are larger for high gravity models than for low gravity
ones. This is because the stronger Lyman-jump and the stronger EUV
flux-blocking (higher densities $\rightarrow$ lower metal ionization stages
$\rightarrow$ more lines) have to be compensated for on the red side of the
flux-maximum to achieve flux conservation. 

If the ionization/recombination rates are dominating, the
(photospheric) departure coefficients inversely scale with the flux at the
corresponding edge (for similar electron temperatures, cf. \citealt{mih78}),
and for higher gravities we obtain lower departure coefficients (more 
ionization) than for lower gravities. Thus, the increase of the
\HeI1.70 line flux with gravity is a final consequence of the different
near UV radiation temperatures as a function of gravity. 

One might wonder why the strength of \HeI4471 is ``well'' behaved, since
this line has the same upper level as \HeI1.70, and the lower level ($2p$
$^{3}P^o$) is strongly coupled to the triplet ground-state as well.
Actually, a simple simulation shows that for this transition the same effect
as for the \HeI1.70 line would be present {\it if the transition were
situated in the IR}. Only because the transition is located in the optical
($h\nu/kT \gg 1$), the corresponding source functions are much less
dependent on gravity (the non-linear response discussed above is largely
suppressed). The profiles react almost only on the opacity,  
which is lower for lower gravity due to the lower number of
available \HeI\ ions.

In summary, the \HeI\ line formation {\it in the optical} is primarily
controlled by different formation regions, since the source functions do not
strongly depend on gravity, whereas {\it in the IR} the deviations from LTE
become decisive. In particular, the influence of considerably different
source-functions is stronger than the different formation depths, where these
source functions are larger for high-gravity models due to a less
overpopulated lower level. 

With respect to the singlet transitions (\HeI2.05, reacting inversely to the
red component of \HeI2.11), we refer the reader to the discussion by
\citet{naj94} and \citet{len04}. But we would like to mention that for a large
range of parameters \HeI2.05 reacts similar to the way described
above, simply because the ionization/recombination rates (over-)populating
the lower level ($2s$ $^{1}{\rm S}$, again a meta-stable level, located at
roughly 3,100~\AA) remain the decisive ingredients
controlling the corresponding source functions. 

Although the basic reaction of \HeI2.05 on gravity is readily
understood from this argument, we like to point out the following (cf.
\citealt{naj94}). For most of the objects considered here, the {\it upper}
level of this transition ($2p$ $^{1}{\rm P}^{\rm o}$) is populated by the
\HeI\ EUV resonance line at 584 \AA\ and other resonance lines (subsequently
decaying to this level). Thus, it strongly depends on a correct description
of line-blocking in this wavelength range, particularly on lines overlapping
with the resonance transitions. Moreover, any effect modifying the EUV will
have a large impact on \HeI 2.05, e.g., wind clumping, if present close to
the photosphere. This makes the line generally risky for classification
purposes and the determination of \teff\ in those stars where \HeII\ is no
longer present (cf.~ Sect.~\ref{general}).

\section{Comparison with results by Lenorzer et al.}
\label{comp}

As already mentioned, \citet{len04} recently presented a first calibration
of the spectral properties of normal OB stars using near infrared lines. The
analysis was based on a grid of 30 line-blanketed unified atmospheres
computed with {\sc cmfgen}. They presented 10 models per luminosity class I, III, and
V, where wind-properties according to the predictions by \citet{vink00} have
been used, and \teff\ ranges from 24,000~K up to 49,000~K (cf. Fig.~\ref
{compcmfgen}). Emphasis was put on the behaviour of
the equivalent widths of the 20 strongest lines of H and He in the {\it J,
H, K} and {\it L} band. For detailed information on procedure and results see
\citet{len04}. In order to check our results obtained by means of {\sc
fastwind}, we have calculated models with identical parameters and
synthesized the same set of NIR lines (see also \citealt{puls05}). Note that
{\sc cmfgen} uses a constant photospheric scale height (in contrast to {\sc
fastwind}), so that the photospheric density structures are somewhat
different, particularly for low gravity models where the influence of the
photospheric line pressure becomes decisive.

\begin{figure}
\resizebox{\hsize}{!}
   {\includegraphics[angle=-90]{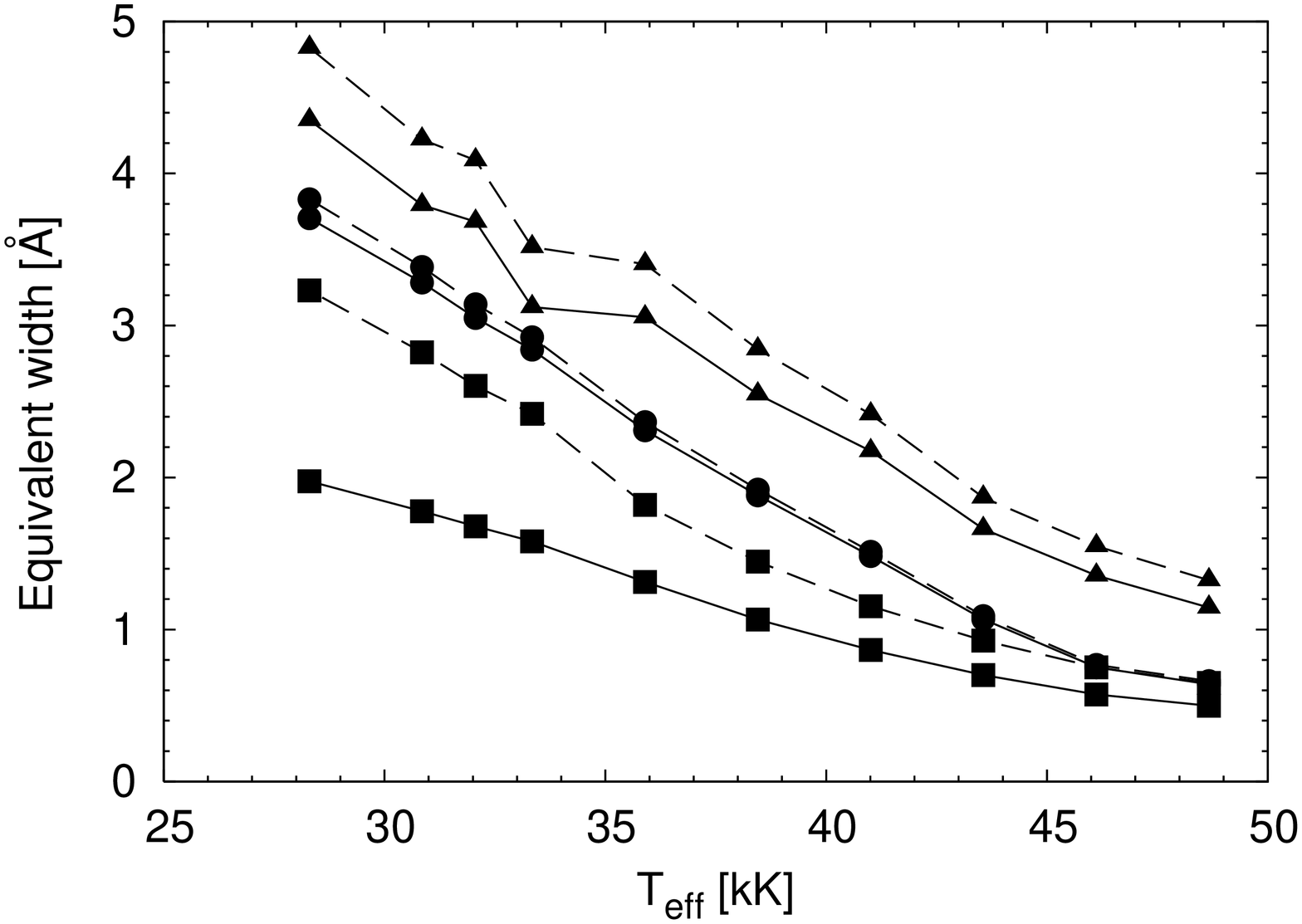}}
\caption{Comparison of equivalent widths (defined here in the conventional
way) for NIR hydrogen lines of O-type dwarfs (\citealt{len04})
using the (erroneous) broadening functions by \citet{lemke97} with results
using broadening functions in the Griem approximation (dashed). 
Squares, circles and triangles correspond to Br11, Br10 and \Brg, respectively.
Note that in all
cases the equivalent width becomes larger and that the most significant
differences occur for Br11 at lower temperatures.}
\label{lemkegriem}
\end{figure}

\begin{figure*}
\begin{center}
\begin{minipage}{16cm}
\resizebox{\hsize}{!}
   {\includegraphics{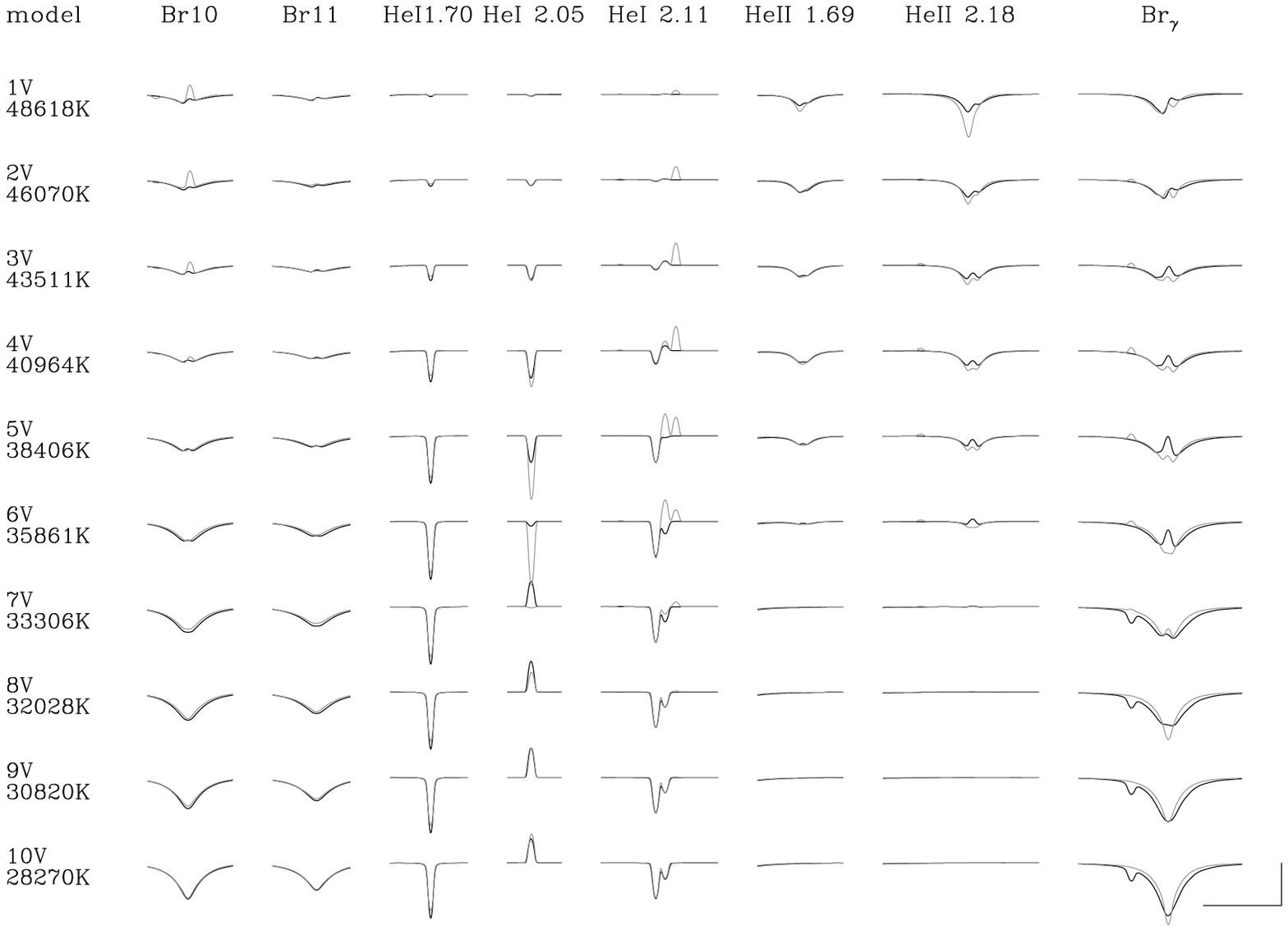}}
\end{minipage}
\hfill
\begin{minipage}{16cm}
\resizebox{\hsize}{!}
   {\includegraphics{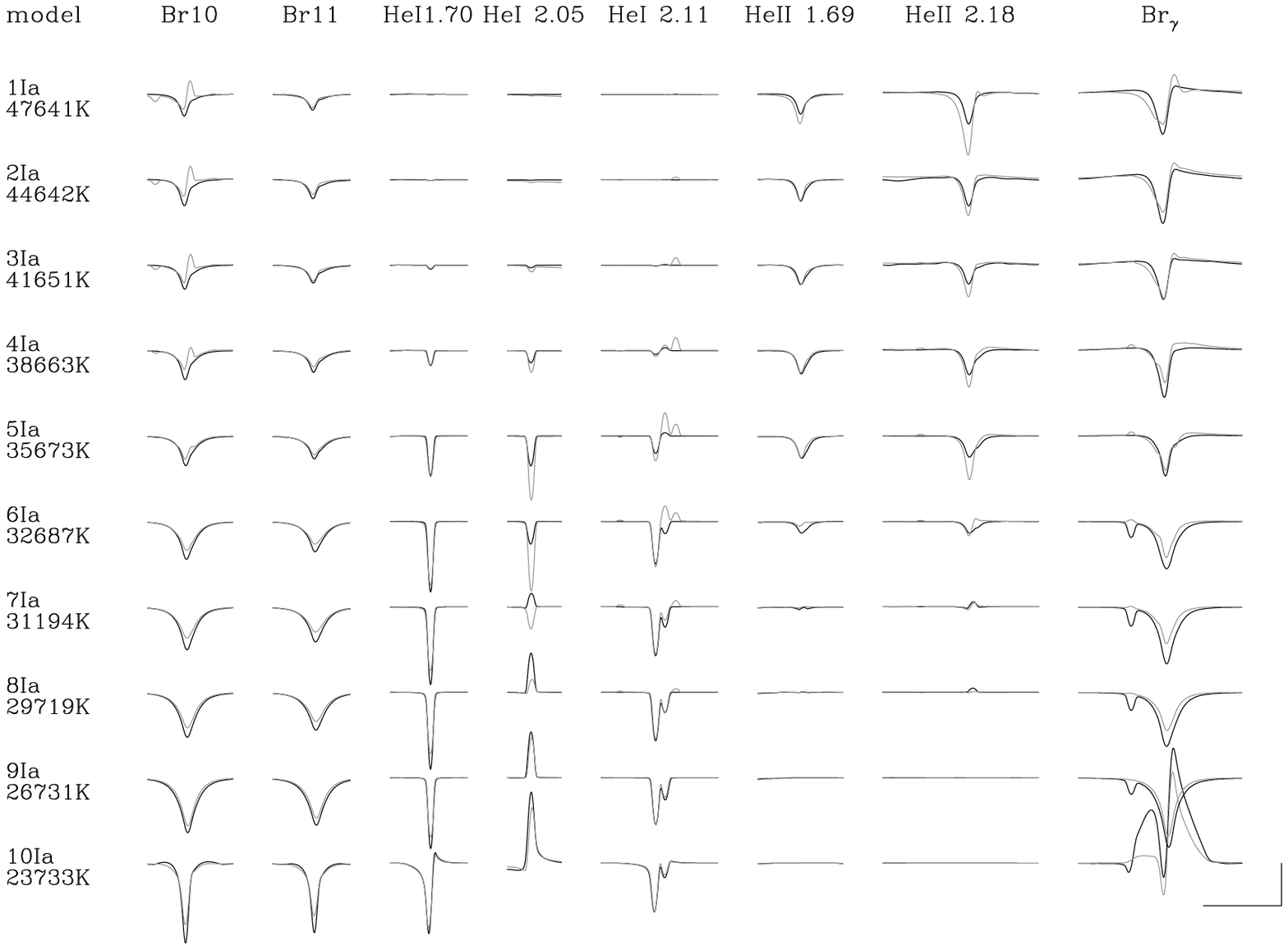}}
\end{minipage}
\end{center}
\caption{Comparison of synthetic NIR lines for the grid of O-type dwarfs
(upper panel) and supergiants (lower panel) described by \citet{len04}, 
as calculated by {\sc fastwind} and {\sc cmfgen} (in grey). 
For hydrogen and \HeII, the results reported by Lenorzer et al. 
have been recalculated using the Griem approximation. The
horizontal and vertical lines in the bottom right corner indicate the scale
used and correspond to 0.01$\mu$m in wavelength and 0.1 in units of the continuum,
respectively. To simplify the comparison, the synthetic profiles have been
convolved with a rotational profile corresponding to $\vsini$ = 80 \kms and
degraded to a typical resolution of 10,000.}
\label{compcmfgen}
\end{figure*}

Since Lenorzer et al. calculated their hydrogenic profiles with the
erroneous broadening functions provided by \citet{lemke97}, the {\it H} and
{\it K} band
profiles have been recalculated by means of the Griem approximation by one
of us (R.M.).  The differences in the equivalent widths for the dwarf grid
are shown in Fig.~\ref{lemkegriem}. In all cases the equivalent widths
became larger, mainly because of increased line wings, and the most
significant changes occurred for Br11 at lower temperatures. Note, however,
that also \Brg\ has become stronger throughout the complete grid.

In Fig.~\ref{compcmfgen} we now compare the detailed profiles of the
strategic lines located in the {\it H} and {\it K} band of the present investigation
(results from {\sc cmfgen} in grey).  The agreement between the results for
the (almost purely photospheric) lines in the {\it H} band (Br10/11, \HeI1.70 and
\HeII1.69) is nearly perfect. The only differences occur in the cores of
Br10, where {\sc cmfgen} predicts some emission for hotter objects, and some
marginal differences in the far wings of the supergiants, which we attribute
to the somewhat different density stratification in the photosphere.
Additionally, {\sc cmfgen} predicts slightly stronger \HeII1.69 lines for the
hottest objects (models ``1V'' and ``1Ia'') and for the supergiant model 
``6Ia''.

Concerning the {\it K} band, the comparison is also rather satisfactory, except
for \HeI2.05 at intermediate spectral type (see below).  Concerning \Brg,
the dwarf models give rather similar results, with the exception of
intermediate spectral types, where {\sc fastwind} produces some central 
emission. We have convinced ourselves that this prediction is very stable
(and not depending on any temperature inversion), resulting from
some intermediate layers where the population of the hydrogen levels is
similar to the nebular case. Here the departure coefficients of the
individual levels increase as a function of quantum number. In such a
situation, a central emission owing to a strong source function is
inevitable. For the supergiants, the major differences regard, again,
the line cores, with {\sc cmfgen} predicting more refilling.

Somewhat larger differences are found for \HeII2.18, again (cf. \HeII1.69) for
the hottest models where {\sc cmfgen} predicts significantly more absorption. 

Concerning \HeI, the situation for the triplet line (blue component of
\HeI2.11) is as perfect as for \HeI1.70. The differences for the \HeI\
(triplet) component located at the blue of \Brg\ are quite interesting. {\sc
cmfgen} predicts an emission for hot stars but ``nothing'' for cooler
objects, whereas {\sc fastwind} predicts a rather strong absorption at
cooler temperatures. To our present knowledge, this is the only discrepancy
we have found so far (including the optical range) 
for a {\it triplet} line. 

The only {\it important} discrepancies concern the \HeI\ singlets (\HeI2.05
and the red component of \HeI2.11) of the supergiant {\it and} dwarf models
in the range between models 5 to 7. Starting from the hotter side,
{\sc cmfgen} predicts strong absorption, which abruptly switches into
emission at models no. 7, whereas {\sc fastwind} predicts a smooth
transition from strong absorption at model 4 to strong emission at model 8.
Reassuring is the fact that at least the inverse behaviour between \HeI2.05
and \HeI2.11(red) (as discussed in \citealt{naj94} and \citealt{len04}) is
always present.

Analogous comparisons performed in the optical (\citealt{puls05}) have
revealed that the strongest discrepancies are found in the same range of
spectral types. The triplets agree perfectly, whereas the singlets disagree,
because they are predicted to be much shallower by {\sc cmfgen} than the
ones resulting from {\sc fastwind}. Again, the transition from shallow to
deep profiles (at late spectral type) occurs abruptly in {\sc cmfgen}. 

\citet{puls05} have discussed a number of possibilities which might be
responsible for the obvious discrepancy, but at present the situation
remains unclear. One might speculate 
that this difference is due to subtle differences in the EUV, 
affecting the \HeI\ resonance lines and thus a number of singlet states, 
as outlined in the previous section.
We will, of course, continue in our effort to clarify this
inconsistency.

\begin{figure}
\resizebox{\hsize}{!}
   {\includegraphics{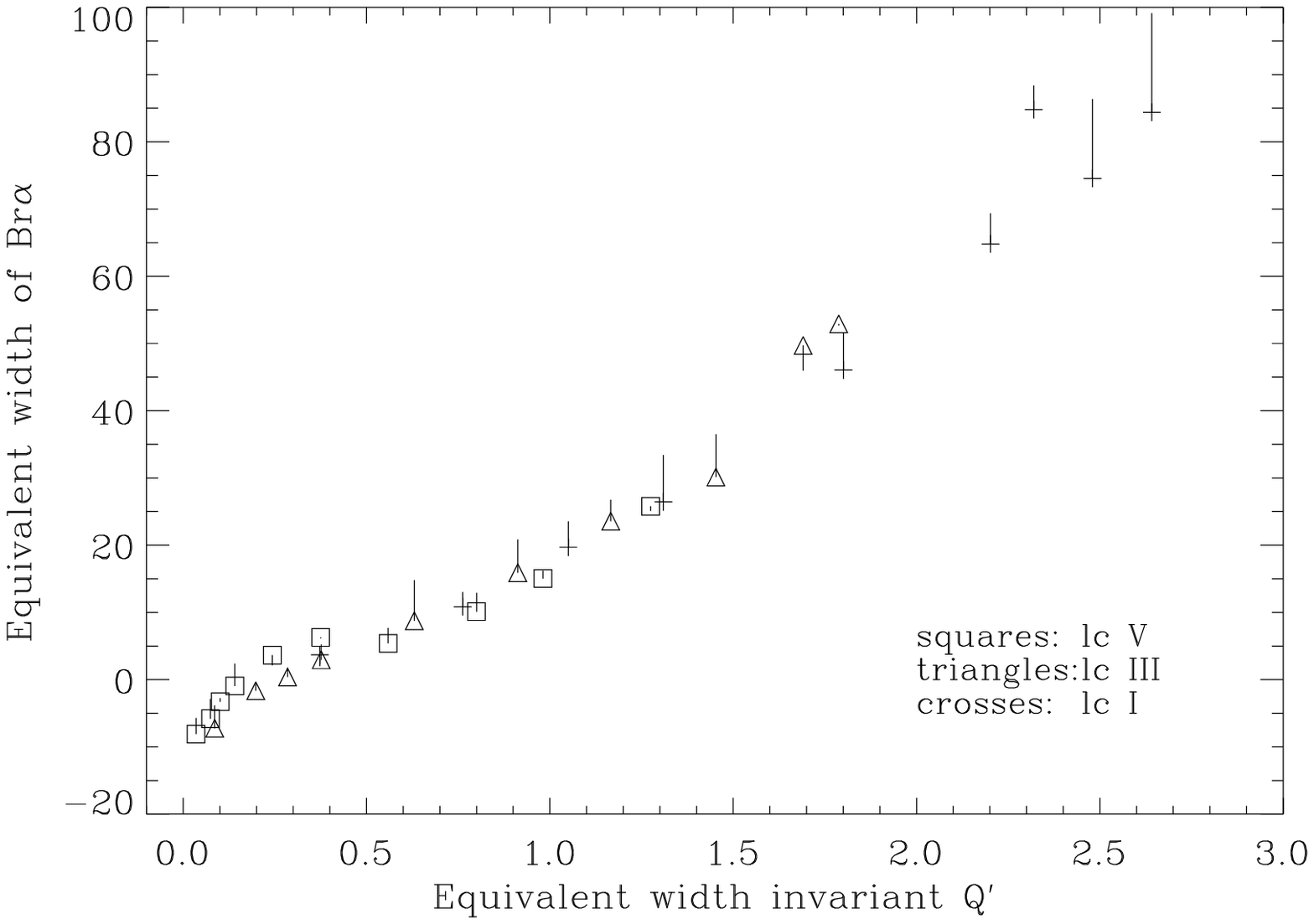}}
\caption{Equivalent width of \Bra\
as a function of equivalent width invariant Q' =
$\frac{\Mdote}{\Rstare^{1.5}\,\Teffe^2\,\vinfe}$ (see \citealt{len04}). 
The dwarf, giant and supergiant models are denoted by
squares, triangles and crosses, respectively. Corresponding values 
from \citet[ their Fig.~7]{len04} are given by the end-points of the vertical
lines (see text).}
\label{bralpha}
\end{figure}

In addition to the detailed comparison performed in the {\it H} and {\it K} band, we have
also compared the resulting e.w.'s of some other important lines in the {\it
J} and {\it L} band. Most important is the comparison for \Bra, which is a primary
indicator of mass-loss, as already discussed in \citet{len04}.
Fig.~\ref{bralpha} compares the corresponding e.w.'s, as a function of
``equivalent width invariant'' $Q'$ (see Lenorzer et al.). Generally, the
comparison is satisfactory, and particularly the differences at large
mass-loss rates are not worrying, since in this range the net-emission
reacts strongly on small changes in \Mdot. Real differences are found only
for the weakest winds, probably related to ``uncorrected'' broadening
functions used by Lenorzer et al.

As we have found for \HeII2.18, also the differences for the other \HeII\ 
lines are significant. Note that we can only compare the e.w.'s and that the
broadening as calculated by Lenorzer et al. suffers from erroneous line
broadening. For \HeII5-7 the behaviour compared to \citet{len04} is the
same, but our lines are twice as strong in absorption in the case of giants
and dwarfs. For the supergiants we obtain the strongest absorption at 36 kK,
in contrast to 42 kK in the comparison models. The only difference found for
\HeII6-11 concerns the behaviour of supergiant and dwarf line trends. The
supergiants in the comparison models show stronger absorption lines than the
dwarfs, whereas in our case the situation is reversed.

At high \teff, the models for \HeII7-13 display a monotonic
behaviour with the hottest models showing the strongest absorption. Our
hottest models display weaker absorption profiles (as was found in the
detailed comparison of \HeII1.69 (7-12)), partly due to
emission in the line cores. Finally, our emission lines obtained for
\HeII6-7 are twice as strong in the case of supergiants and giants compared
to \citet{len04}.

In summary we conclude that at least from a theoretical point of view, all
{\it H/K} band lines synthesized by {\sc fastwind} can be trusted, except
for \HeI2.05 at intermediate spectral type and maybe \HeII2.18, where certain
discrepancies are found in comparison with {\sc cmfgen}, mostly at hottest
temperatures. Concerning the discrepancies of \HeII\ in other bands, we have
to clarify the influence of correct broadening functions, whereas for the
\HeI\ singlet problem work is already in progress. 

\section{Analysis}
\label{analysis}

\subsection{General remarks}
\label{general}

It might be questioned to what extent all decisive stellar and wind parameters can be
obtained from a lone IR-analysis in the {\it H} and {\it K} band. In view of the
available number of strategic lines, however, in most cases we are able to
obtain the full parameter set, except for 
\begin{enumerate}
\item[i)] the terminal velocity, which in most cases cannot be derived from
the optical either, and has been taken from 
UV-measurements. For our analysis, we have used the values given in the 
publications corresponding to sample~I to III. The terminal velocities of
sample IV have been adopted from \citet{howarth97}.
If no information is (or will be) present, calibrations of
\Vi\ as a function of spectral type have to be used, e.g., \citet{kupu00}.
\item[ii)] the stellar radius, which can be inferred from \MV\ and the
theoretical fluxes (\citealt{kud80}), and has been taken from the optical
analyses in the present work. In future investigations when no optical data
will be available, a similar strategy exploiting infrared colors can
certainly be established. 
\end{enumerate}
In particular, Br10 and Br11 give clues on the gravity (if \teff\ is known),
\HeI\ and \HeII\ define temperature and helium content, and \Brg\ can serve as
an \Mdot\ indicator, at least in principle. In those cases, where only
one ionization stage of helium is visible, the determination of
$Y_{\rm He}$ becomes problematic, and also the uncertainty for \teff\
increases (see below). Due to the high quality of our spectra, however,
both \HeII\ lines are visible for most spectral types. 

Only for the coolest objects \HeII\ vanishes, which occurs for spectral
types later than O9 for dwarfs, about B0 for giants, and again B0 for
supergiants (cf. Figs.~\ref{dwarfs} to \ref{supergiants}). In those cases 
it still should be possible to derive (somewhat more inaccurate) estimates
for \teff, at least if some guess for $Y_{\rm He}$ is present. This
possibility is due to the behaviour of \HeI1.70 vs. \HeI2.05 
(Fig.~\ref{isocontours}), since the former line is almost only dependent on
\g, whereas the latter depends strongly on \teff\ (with all the caveats
given earlier on). Unfortunately,
the data for \HeI2.05 are not of sufficient quality (except for HD\,190864 and
$\tau$ Sco, where the latter just lies in the critical domain) that we could
exploit this behaviour only once and had to refrain from an analysis
of the remaining coolest objects (four in total). 

Because of the independence of \HeI1.70\ on \teff\ and the fact that Br10/11
can {\it always} be fitted for certain combinations of \teff/\g, a perfect
fit in combination with completely erroneous parameters would result if
\HeI2.05 had to be discarded. This is indicated in Fig.~\ref{supergiants}
for HD\,14134, being a B3Ia supergiant (with \teff\ roughly at 18,000~K, see
\citealt{kudetal99}), which could be fitted with \teff\ = 25,000~K. If, on
the other hand, \HeI2.05 had been available, the appropriate parameters
should have been obtained, at least when the helium content could have been
guessed. Such a guess of the helium abundance should always be possible for
objects we are eventually aiming at in our project (cf. Sect.~\ref{intro}),
i.e., for very young, un-evolved stars with unprocessed helium.

\paragraph{Micro-turbulence.} In agreement with the findings 
by \citet{repo04}, we have adopted a micro-turbulence of \vturb\ =
10\kms\ for all stars with spectral type O7 or later regardless of their
luminosity class, whereas for hotter O-type stars the micro-turbulent
velocity has almost no effect on the analysis and we have neglected it.  At
spectral type O6.5, our IR-analysis of HD\,190864 (O6.5 III) indicated that a
micro-turbulence is still needed, whereas from O7 onwards \vturb\ did not
play any role, e.g., for HD\,192639 (O7 Ib). Since the former and the latter
stars have \teff\ = 37 and 35 kK, respectively, we conclude that at roughly 
\teff\ = 36 kK the influence of \vturb\ on the H/He lines is vanishing, in
agreement with our previous findings from the optical.

\paragraph{Rotational velocities.} For the (projected) rotational
velocities, we have, as a first guess, used the values provided by 
\citet{repo04}, \citet{h00, h02} and \citet{howarth97} for sample~I, II and
III/IV, respectively. In our spirit to rely on IR data alone, we have
subsequently inferred the rotational velocity from the (narrow) He
lines, with most emphasis on \HeI. Concerning sample~I, the results from our
IR-analysis are consistent with the velocities derived from the optical,
except for HD\,190864 and HD\,192639, where the profiles
indicated slightly lower values (10\% and 20\%, respectively), which have
been used instead of the ``original'' ones. 

For sample~II stars, in 3 out of 5 cases the ``optical'' values derived by
\citet{h00, h02} were inconsistent with our IR-data. In particular, for
HD\,5689 we found a velocity of 220 \kms\ (instead of 250 \kms), for
HD\,15570 a velocity of 120 \kms\ (instead of 105 \kms) and for Cyg\,OB2\#7
our analysis produced the largest differences, namely \Vr\ = 145 \kms,
compared to a value of 105 \kms provided by \citet{h02} (30\% difference!). 

The values taken from \citet{howarth97} for the remaining sample~III/IV 
objects, finally, agree fairly well with our IR data, and are also
consistent with the values derived by \citet{kudetal99} in their analysis of
sample~III objects.

\smallskip
\noindent
Let us finally mention that in those cases when \Brg\ does not show emission 
wings, a statement concerning the velocity field parameter, $\beta$, is
not possible, as is true for the optical analysis. In order to allow for
a meaningful comparison with respect to optical determinations of \Mdot,
we have used the corresponding values derived or adopted
from the optical. In future analyses, of course, this possibility will no
longer be present, and we have to rely on our accumulated knowledge, i.e.,
we will have to adopt ``reasonable'' values for $\beta$, with all related
problems concerning the accuracy of \Mdot\ (cf. \citealt{puls96, markova04}).

\subsection{Fitting strategy and line trends} 
\label{fitstrat}

In order to obtain reliable fits, we applied the following strategy. At
first, we searched for a coarse determination of the relevant sub-volume in
parameter space by comparing the observed profiles with our large grid of
synthetic profiles as described by \citet{puls05}, which has a typical
resolution of 2,500~K in \teff, 0.3 in \g, and 0.25 in $\log Q$. A
subsequent fine fit is obtained by modifying the parameters by hand (using
the ``actual'' values for \Rstar\ and \Vi\ to obtain information on \Mdot\
additionally to $\log Q$), where typically 10 trials are enough to provide a
best compromise. In those cases, where at present no information about
\Rstar\ is available (which concerns the three objects presented in
Table~\ref{paranew}), ``only'' $\log Q$ can be derived. For the actual fits
of these three objects we have, of course, used prototypical parameters for \Rstar\ and
\Vi. Further discussion of related uncertainties is given in
Sect.~\ref{logqapproach}.

Most weight has been given to the fits of the He lines (which are rather 
uncontaminated from errors in both broadening functions and reduction of the
observed material) followed by the photospheric hydrogen lines, Br10/11, 
which sometimes strongly suffer from both defects. Least weight has been
given to \Brg, because of the number of problems inherent to this line, as recently
described by \citet{len04} and independently found by
\citet{jok02}. Particularly, the synthetic profiles for larger wind
densities, predicted by both {\sc fastwind} and {\sc cmfgen}, are of P~Cygni
type, whereas the observations show an almost pure emission profile.
Moreover, from a comparison of equivalent widths, it has turned out that in
a lot of cases the predicted e.w. is much larger than the observed
one, which would indicate that the models underestimate the wind-emission
(remember, that \Brg\ forms inside the \Ha\ sphere). Often, however,
this larger e.w. is due to the predicted P~Cygni absorption component which
is missing in the observations, and we tried to concentrate on the \Brg\
line wings in our fits ignoring any discrepancy concerning the predicted 
P~Cygni troughs. If the synthetic lines actually predicted too few wind 
emission, this problem would become severe for lines where pure
absorption lines are observed, and should lead to an overestimate of \Mdot.
We will come back to this point in the discussion of our analysis.

Another important point to make concerns the \HeI\ \lam 2.11 line
(comprising the \HeI\ triplet \lam 2.1120 and the \HeI\ singlet \lam
2.1132). Close to its central frequency, a broad emission feature can be
seen (at \lam 2.115) in the spectra of hot stars. This line can either be
identified as \NIII\ (n = 7 \rarrow 8) or as \CIII\ (n = 7 \rarrow 8) or
maybe both (\citealt{hanson96, najarro97, naj04}).\footnote{Due to the
rather similar structure and the fact that these transitions occur between
high lying levels, the predicted transition frequencies are almost equal.
Since most of the stars in the OIf phase will have depleted C and enhanced
N, however, the major contribution should be due to \NIII\ and possibly also
due to \NV2.105 (10 \rarrow 11) for the hottest objects (F. Najarro, priv. 
comm.). \CIII\ will be contributing if \CIV\ at 2.07-2.08$\mu$m is strong.} 
This feature is seen in stars of all luminosity classes, for stars
hotter than and including spectral type O8 in the case of dwarfs and giants
and O9 in the case of supergiants (though its 
designation is somewhat unclear, as \HeI2.11 resembles a P Cygni profile in
late-O supergiants, possibly mimicking this feature). Since our present
version of {\sc fastwind} synthesizes ``only'' H/He lines and their analysis
is the scope of the present paper, we are not able to fit this feature, but
have to consider the fact that this feature significantly contaminates
\HeI2.11.

Due to the well-resolved spectra, the two \HeI\ lines overlapping with \Brg\
as mentioned in Sect.~\ref{obs}, i.e., the \HeI\ triplet \lam 2.1607 and the
\HeI\ singlet \lam 2.1617, are also visible in certain domains. For
supergiants later than O5, \HeI2.1607 begins to appear in the blue wings of
\Brg, and in two stars, HD\,30614 and HD\,37128, the \HeI2.1617 singlet
seems to be present, even if difficult to see. In the giant spectra,
\HeI2.1607 can be seen from spectral type O9 onwards, and in the dwarf
spectra this line appears in spectral types later than O8. 

The strength of the Brackett lines in supergiants (Fig.~\ref{supergiants}) 
shows a smooth behaviour as a function of spectral type, apart from certain
fluctuations such as blends in the late O- and early B-type stars. As one
moves from early B-type to mid O-type (i.e., O5), the \Brg\ absorption
weakens, and from mid to the earliest O-types the line profiles switch into
emission, where the emission at the blue wings of \Brg\ is much more
pronounced (except for HD\,15570), presumably due to the overlapping \HeII\
blend. 

As for the photospheric Br10/11 lines, we can see that these absorption
profiles show an extremely continuous behaviour, being rather weak for early
O-type stars and increasing in strength towards early B-types. Hence, the 
cooler supergiants show the most prominent and sharpest
absorption features. The emission features visible at the blue side of Br10 in
the hottest supergiants are due to an unidentified feature.

Fig.~\ref{supergiants} shows that the observed Br10/11 profiles are mostly
well reproduced by the theoretical predictions, although at hotter
temperatures certain inconsistencies arise, particularly with respect to the
line cores. Most interestingly, in a number of cases we could not fit both
profiles in parallel, and typically Br11 is then of better quality.  Since
we have convinced ourselves that the differences most probably are not
exclusively due to reduction problems, we repeat our hypothesis that the
broadening functions are somewhat erroneous, cf. Sect.~ \ref{stark}. Again,
for the theoretical profiles for \Brg, we would like to mention that for
emission lines the wings are fairly well reproduced in contrast to the line
cores.

The \HeI1.70 line shows a very smooth behaviour, being absent in the
hottest and most luminous star, Cyg\,OB2 $\#$7, and successively increasing 
towards late O-type and early B-type stars. This also applies to the
sharpness of the profiles. As has been stressed earlier on, \HeII1.69 and
\HeII2.18 vanish in supergiants of spectral type B0 (being still detectable for
$\alpha$ Cam, O9.5Ia)
 
\begin{figure*}
\begin{center}
\begin{minipage}{16cm}
\resizebox{\hsize}{!}
   {\includegraphics{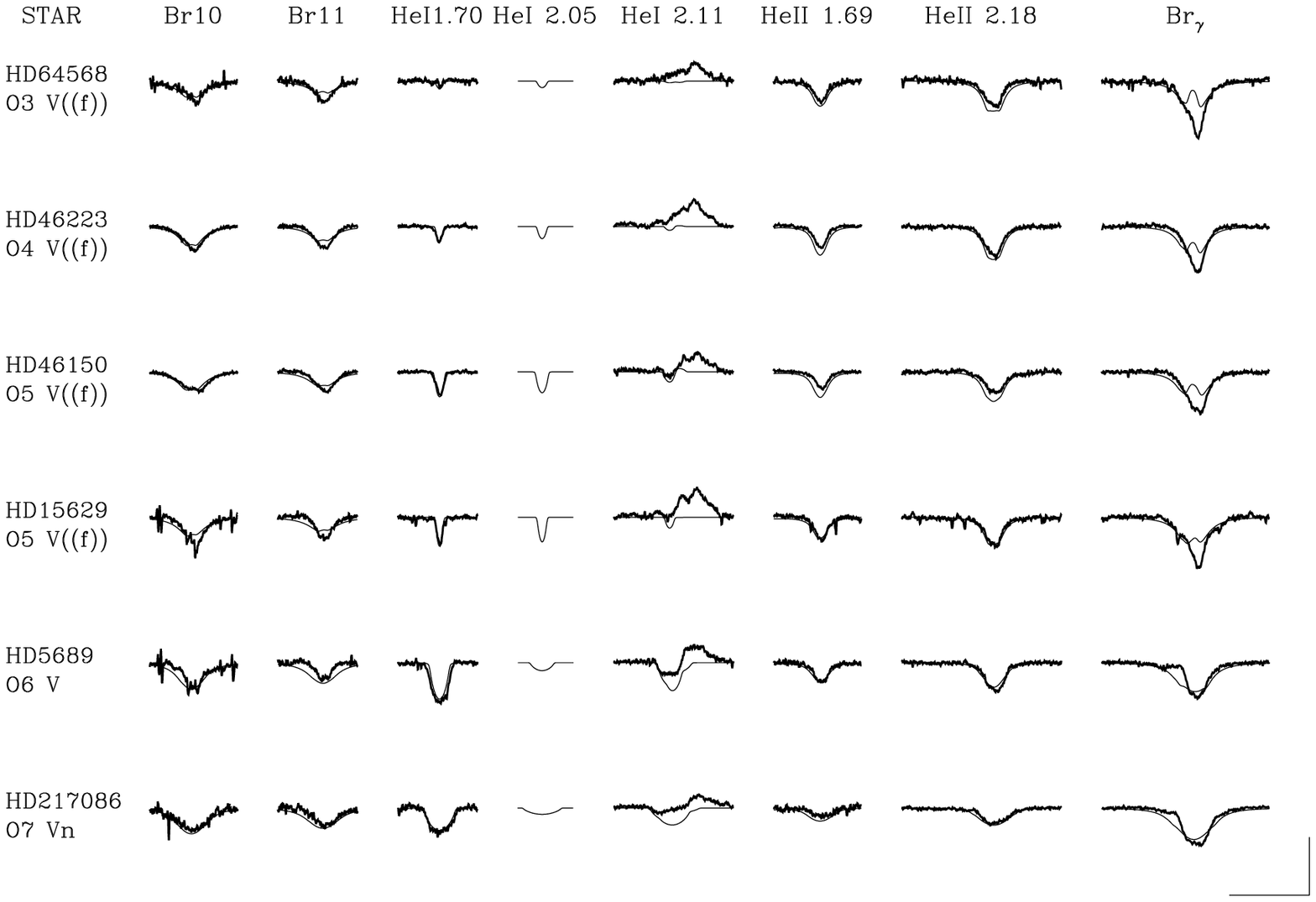}}
\end{minipage}
\hfill
\begin{minipage}{16cm}
\resizebox{\hsize}{!}
   {\includegraphics{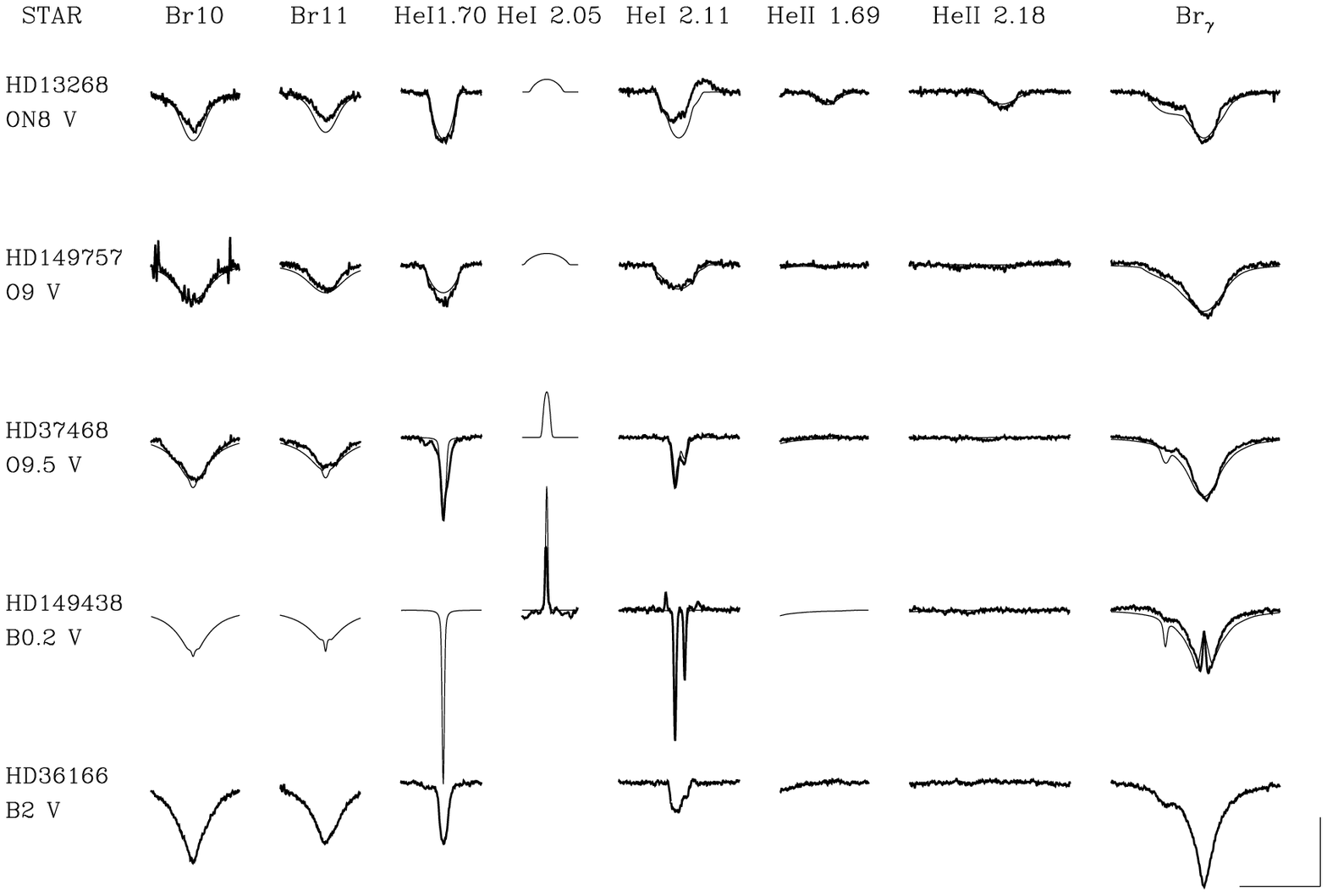}}
\end{minipage}
\end{center}
\caption {Line fits for hot dwarfs with spectral types ranging from O3 to O7
(upper panel) and cool dwarfs with spectral types ranging from O8 to B2
(lower panel). The
lowermost object (HD\,36166, B2V) has not been analyzed due to missing \HeII\ and
\HeI2.05 (see text). The horizontal and vertical lines in the bottom right
corner indicate the scale used and correspond to 0.01 microns in wavelength
and 0.10 in units of the continuum, respectively.}
\label{dwarfs}
\end{figure*}

The situation is similar in the case of giants (Fig.~\ref{allgiants}) and 
dwarfs (Fig.~\ref{dwarfs}). All hydrogen and \HeI\ lines show the systematic
variations expected, namely, an increase in strength from early O-types to
early B-types.  The model predictions do agree well with the observed
profiles, again, except for certain discrepancies between Br10 vs. Br 11.
Since \Brg\ remains in absorption throughout the entire spectral range, it
can be reasonably fitted in most cases (whether at the ``correct'' value, will
be clarified in Sect.~\ref{compdata}). Particularly, the \HeII\ profiles
give almost perfect fits except for very few outliers, and vanish at 09
for dwarfs and about B0 for giants.

\subsection{Comments on the individual objects}

In the following, we will comment on the fits for the individual objects
where necessary. Further information on the objects can be found in the
corresponding publications concerning the optical analyses, see
Tab.~\ref{runs}. A summary of all derived values can be found in
Tables~\ref{paranew} and \ref{paraold}.

\subsubsection{Dwarfs}
\label{dwarfcomments}

\begin{figure*}
\resizebox{\hsize}{!}
   {\includegraphics{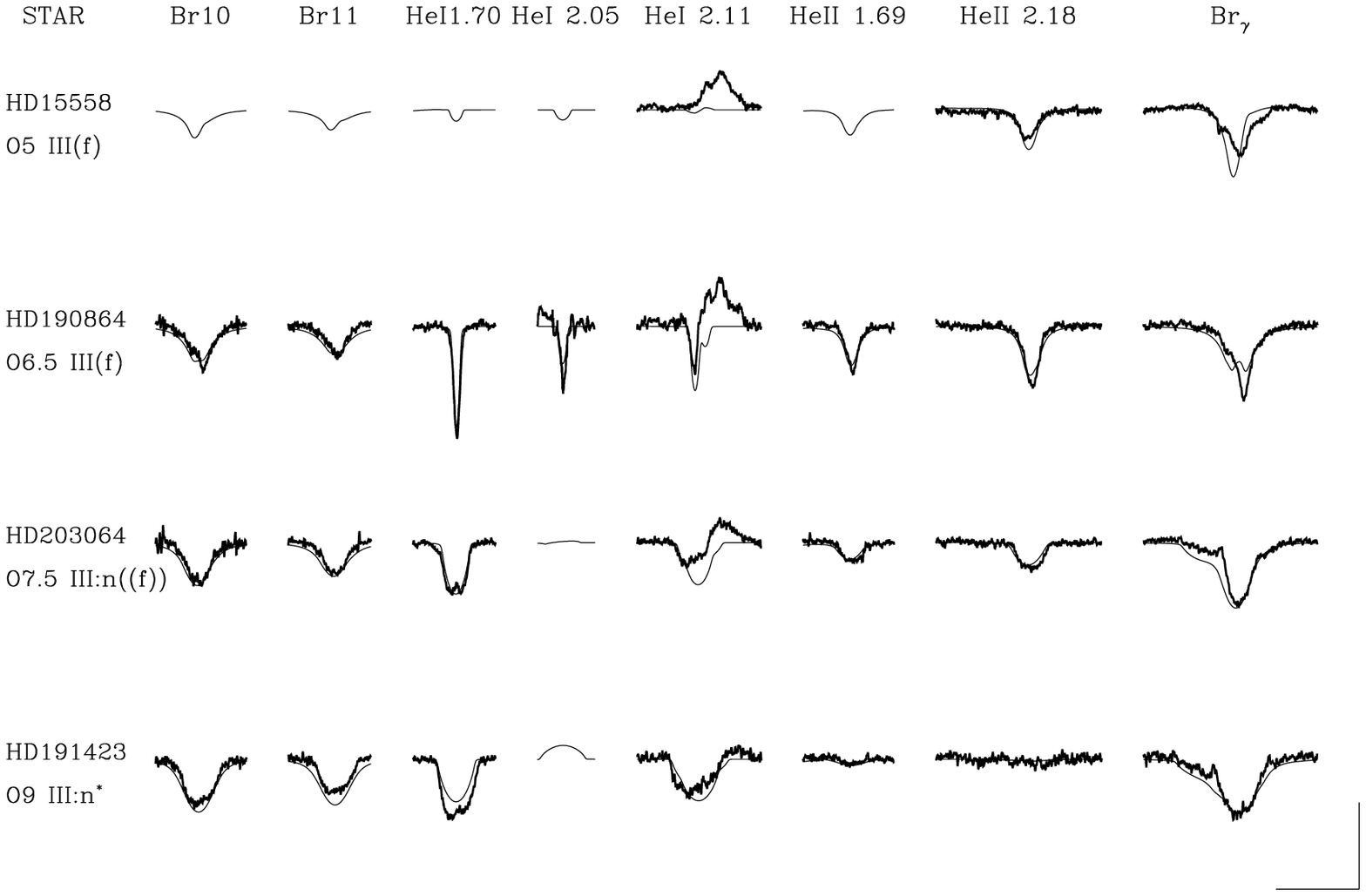}}
\caption{As Fig.~\ref{dwarfs}, but for giants with spectral 
types ranging from O5 to B9. 
Concerning HD\,15558 (only {\it H} band available), see text.}
\label{allgiants}
\end{figure*}

\paragraph{HD\,64568.} The fit quality of the lines is generally good except
for \Brg. The theoretical profile displays a central emission which
is more due to an overpopulated upper level than due to wind effects (cf.
Sect.~\ref{comp}) and, thus, cannot be removed by adopting a lower \Mdot. 
Moreover, the theoretical \HeII\ line would become too strong if a lower
\Mdot\ were used. Insofar, the present fits display the best compromise.
Since no radius information is available, only the optical depth invariant
$\log Q$ is presented.  

\paragraph{HD\,46223 and HD\,46150.} For both objects, the fit quality is
satisfactory, except for the \HeII\ lines (particularly in HD\,46150) and \Brg.
The former lines are predicted to be too strong and for the latter
there is, again, too much central emission present. It is, however, not
possible to reduce the temperature in order to fit the \HeII\ line, since
this would adversely affect \HeI.  A further reduction of \Y\ (adopted here
to be ``solar'', i.e., 0.1) is implausible, so that the presented line fits
reveal the best fit quality possible. For HD\,46223 we can derive only the
optical depth invariant $\log Q$ due to missing radius information.

\paragraph{HD\,15629.} The fit quality for the He lines is very
good, and we confirm the helium deficiency to be \Y\ = 0.08 as determined in
the optical, see \citet{repo04}. \Brg, again, suffers from too much central
emission, and the cores of Br10/11 are much narrower than predicted (at least
partly, as some of the narrowness might be due to reduction problems).
The mass-loss rate is moderate with a value of $\Mdote \approx$ 1.3~\Mdu\
which represents the same value as determined in the optical, whereas 
\g\ is found to be larger by 0.1 dex.

\paragraph{HD\,5689.} Again, moderate mismatches for the H lines are found,
whereas the He lines provide a good fit. \Brg\ does not show a central
emission anymore, but the theoretical profile seems to be too broad. The
same problem (very steep increase on the blue side, almost perfect fit on
the red side) seems to be present also in HD\,217086 (and, to a lesser
extent, in HD\,203064 and HD\,191423), and we attribute some of this 
disagreement to reduction errors, although
an underestimate of the \HeII\ blend (which is in emission in this parameter
range) might be possible as well. Since all four stars are very fast
rotators, effects from differential rotation in combination with a
non-spherical wind (cf. \citealt{puls96, repo04}
and references therein) cannot be excluded, see below.
 
In the case of Br10/11, on the other hand, problems in the broadening
functions might explain the disagreement, as already discussed. Finally,
the absorption trough of the theoretical profile for \HeI2.11\ seems to be
too strong, but might be contaminated by the bluewards \NIII/\CIII\
complex.

\paragraph{HD\,217086.} A very similar fit quality as found for HD\,5689 has been
obtained for this star, although Br10/11 are now in better agreement.
The parameters determined are comparable to the ones obtained from the
optical, including the overabundance of He (\Y\ = 0.15). 
An upper limit for the mass-loss rate has been derived, which is 
less than half the value obtained from the optical. 

\paragraph{HD\,13268.} The theoretical prediction reproduces the observation
quite well, especially in the case of \HeI1.70 and both \HeII\ lines. As for the
hydrogen lines, the two photospheric lines Br10 and Br11 show too much
absorption in the line cores, whereas \HeI2.11\ shows the same trend as
already discussed for HD\,5689. The fit quality for \Brg, however, is much
better, and even the \HeI2.1607 (triplet) blend is reasonably reproduced,
although slightly too strong. For the mass-loss rate only an upper limit of
0.17 \Mdu\ can be given, for an adopted value of $\beta$ = 0.80. The
enhanced helium abundance \Y\ = 0.25, as found in the optical, could be
confirmed, giving the best compromise regarding all He lines.

\paragraph{HD\,149757 and HD\,37468.} 
The very good fit quality makes further comments unnecessary.

\paragraph{HD\,149438.} $\tau$ Sco is probably one of the most interesting
stars of the sample analyzed, since it is a {\it very} slow rotator and all
features become visible at the obtained resolution. Although only the {\it K} band
observation is available, it can be seen that we obtain a very good fit
quality for all H and He lines present (\HeII\ is absent at these
temperatures). As discussed before, in those cases where only one ionization
stage of helium is visible, the determination of $Y_{\rm He}$ becomes
problematic, and also the uncertainty for \teff\ increases. Since in the
case of $\tau$ Sco we could make use of the \HeI2.05 line, we could still
determine the effective temperature (resulting in a similar value as in the
optical), on the basis of an adopted value \Y\ = 0.1. Also the mass-loss rate
is well constrained from the {\it resolved} central emission feature in
\Brg, having a value of 0.02 \Mdu. From a similar investigation by
\citet{PB04}, exploiting the central emissions of \Pfg, \Pfb\ and \Bra\ as
well, they derived a value of 0.009 \Mdu\ (factor two lower) as a
compromise, but have adopted a different velocity-field exponent ($\beta =
2.4$ instead of $\beta = 1.0$ used here) and utilized the ``canonical''
value for \g\ = 4.25 which fits \Hg. In our case, however, and in the spirit
to rely on a lone IR analysis, we preferred a lower value, \g\ = 4.0,
since in this case the emission feature is better reproduced (much narrower)
than for a higher gravity, whereas the differences in \HeI2.05 (and
concerning the line wings of \Brg!) are almost negligible. If we have had
the information on Br10/11, this dichotomy could have been solved.

Having finished our investigation, one of us (R.M.) has analyzed the {\it
optical} spectrum of $\tau$ Sco, also by means of {\sc fastwind}. Details
will be published elsewhere (\citealt{mokiem05}). Most interestingly, he
obtained perfect line fits, at parameters \teff = 31,900~K, \g\ = 4.15,
\Y\ = 0.12 and \Mdot\ = 0.02 {\ldots} 0.06~\Mdu\ (for velocity exponents
$\beta$ = 2.4 {\ldots} 0.8, respectively). We like to stress that this
analysis has {\it not} been biased by our present results from the IR, since
it was performed by an ``automatic'' line fitting procedure based on a
genetic algorithm. 

\paragraph{HD\,36166.} This object has not been analyzed, due to missing
\HeII\ {\it and} \HeI2.05 lines.

\subsubsection{Giants}

\paragraph{HD\,15558.} Also for this star, only the {\it K} band observation is
available, and because of the high temperature and rather large \Mdot\ no
independent information concerning \teff\ and \g\ can be derived. Thus, we 
adopted the effective temperature at its ``optical'' value, \teff\ =
41,000~K. With this value, a simultaneous ``fit'' of \g, \Y\ and \Mdot\
resulted in the synthetic spectrum displayed. \Mdot\ was constrained by the
wings of \Brg, and \Y\ = 0.08 derived on the basis that at this value \HeII\
is still somewhat too strong. \g\ is rather badly defined, since a variation
by $\pm 0.2$ dex gives only small differences in all three observed lines.
In conclusion, the fit obtained allows to reliably constrain the mass-loss
rate alone, and this only {\it if} the temperature actually has the adopted
value. Note, however, that a (much) lower value is excluded since the
predicted \HeII2.18 line would become too weak (cf. Fig~\ref{isocontours},
lower right panel).

\paragraph{HD\,190864.} The analysis gives a consistent fit for all lines
(including \HeI2.05!) except for \Brg, where the theoretical profile of
\Brg\ shows too much central emission.  The parameters remained almost the
same compared to the optical except for the helium abundance, $Y_{\rm He}$,
which has been increased from 0.15 to 0.20.

\paragraph{HD\,203064 and HD\,191423.} The analysis for HD\,203064 yields a
consistent fit for all lines, except for \HeI2.11 which displays a similar
problem as described for HD\,5689. We recovered the same values for \teff\ and
\g\ as in the optical, though the helium abundance had to be doubled and
also the mass-loss rate increased by roughly 80\%. The theoretical profile
of \Brg\ for both stars is slightly broader than observed, although the
effect is weaker than found for HD\,5689 and HD\,217086. Note in particular
that for both giants \Ha\ turned out to be narrower than predicted,
with ``emission humps'' present on both sides of the absorption trough
(\citealt{repo04}, Fig.~6). Summarizing and considering their extreme rotation
velocities (\Vr\ being 300 \kms and 400 \kms for HD\,203064 and HD\,191423,
respectively), our above hypothesis of rotational distortion is the most
probable solution for the apparent dilemma in these cases.

Also for all other lines, HD\,191423 behaves very similarly 
to HD\,203064, although a better fit quality for \HeI2.11 is found, while
\HeI1.70 has become worse (we aimed for a compromise between both lines).

\subsubsection{Supergiants}
\label{sgcomments}

\begin{figure*}
\begin{center}
\begin{minipage}{16cm}
\resizebox{\hsize}{!}
   {\includegraphics{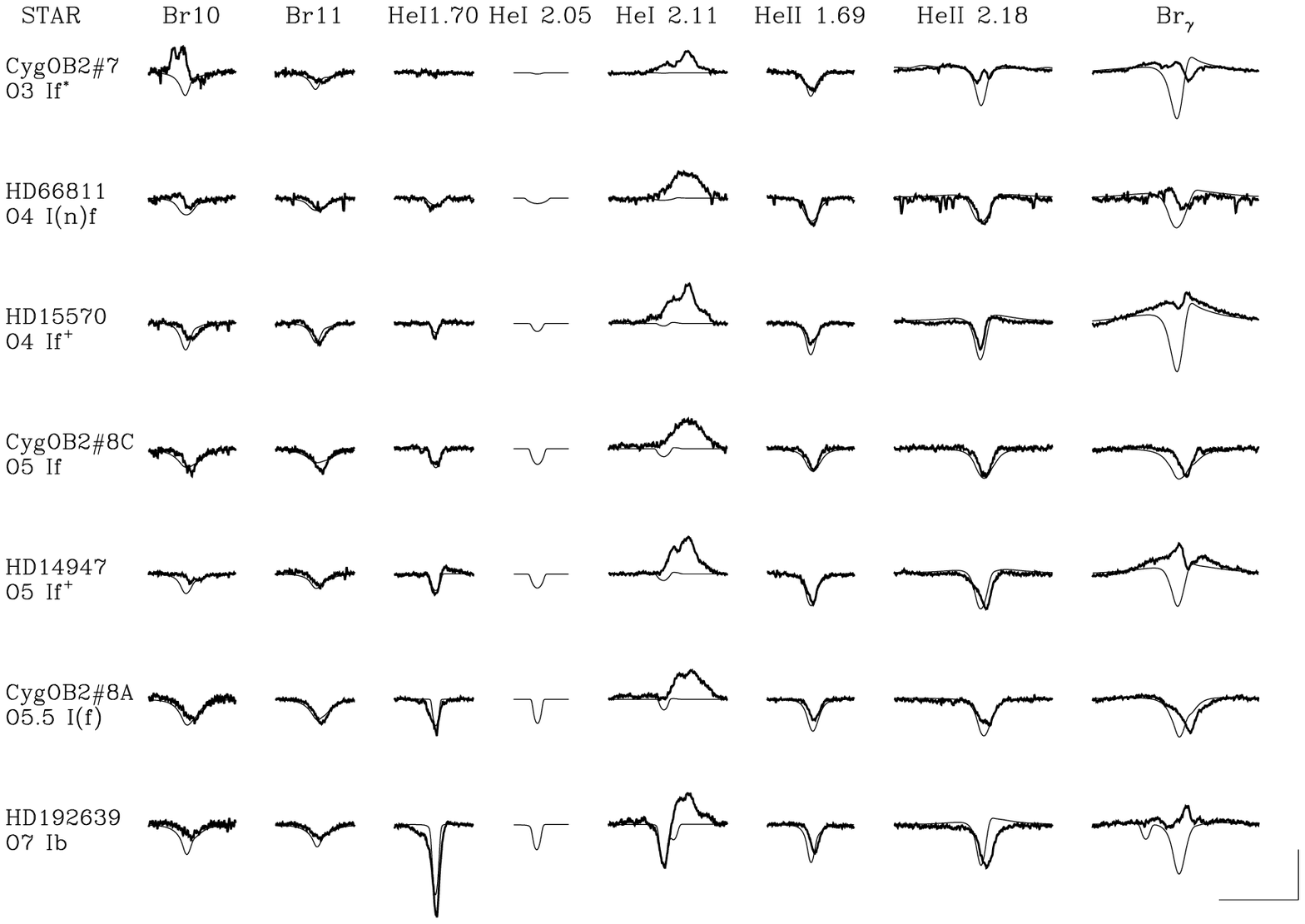}}
\end{minipage}
\hfill
\begin{minipage}{16cm}
\resizebox{\hsize}{!}
   {\includegraphics{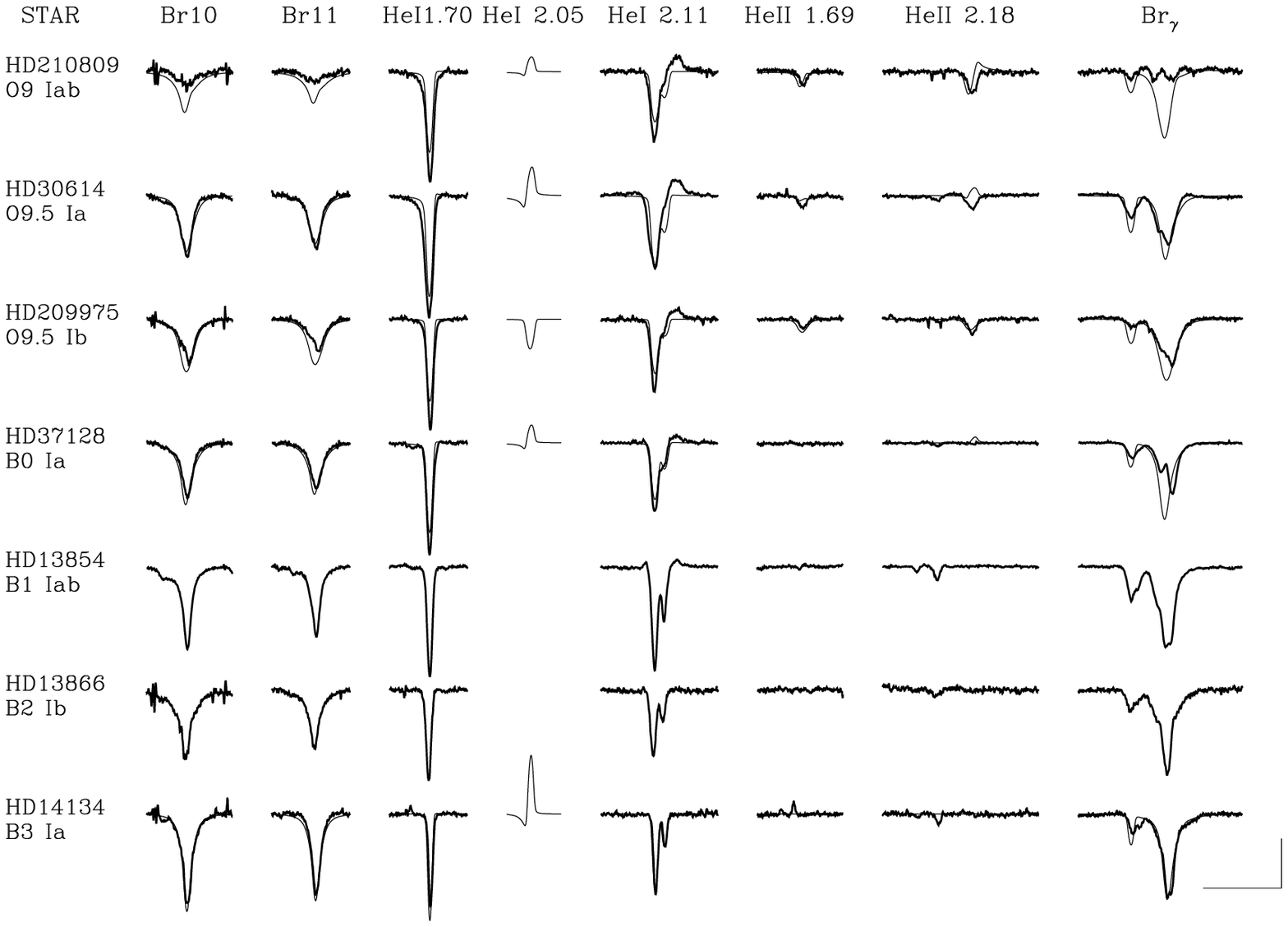}}
\end{minipage}
\end{center}
\caption{As Fig.~\ref{dwarfs}, but for hot supergiants of spectral type 
O3 to O7 (upper panel) and cool supergiants (O9 to B3, lower panel). The three
lowermost objects (HD\,13854, B1Iab, HD\,13866, B2Ib and HD\,14134, B3Ia) 
have not been analyzed. The synthetic profiles overplotted for HD\,14134 show
a perfect fit for completely wrong parameters (\teff = 25,000~K, \g =
2.7), indicating that a spectroscopic H/K band analysis is impossible if
\HeII\ and/or \HeI2.05 are missing (see text).}
\label{supergiants}
\end{figure*}

\paragraph{Cyg\,OB2 $\#$7.} This star, being the hottest one in the sample,
shows an enormous discrepancy in the \Brg\ line, due to the observed central
emission, which is not predicted by our simulations. It is the only star in
our sample where we find the same problem in \HeII2.18, i.e., where the
theoretical predictions with respect to its morphology could not be
confirmed.  In order to determine a fairly reliable mass-loss rate, we have
concentrated exclusively on the wings of \Brg. The parameters derived agree 
with their values from the optical, except for the helium abundance. The
determination of this quantity is problematic due to missing \HeI. In
contrast to the optical value, \Y\ = 0.3 (\citealt{h02}), our best fit
favoured \Y\ = 0.1, whereas simulations using the optical value have
adversely affected the H lines. Moreover, to preserve the good fit quality
of \HeII1.69, we would have to lower \teff\ significantly if \Y\ = 0.3 were
the correct value. (Actually, a temperature already lower by 1,500~K
compared to the optical has been used to achieve the displayed fit).
Interestingly, a re-analysis of Cyg\,OB2 $\#$7 in the optical
performed by one of us (R.M.) resulted in a value just in between, namely 
\Y\ = 0.21 (at \teff\ = 46,000~K). 
The emission on the blue side of Br10 is due to an unknown feature, as discussed
in Sect.~\ref{fitstrat}.
 
\paragraph{HD\,66811.} The fit quality is generally good, except for 
\Brg, which again shows much more central emission than predicted. 
The wings, on the other hand, could be well fitted and 
gave a mass-loss rate of 8.8 \Mdu, in agreement with the 
optical value. Br10 is contaminated by an unknown feature on the blue 
side, but to a lesser extent than in Cyg\,OB2 $\#$7.

\paragraph{HD\,15570 and HD\,14947} show very similar profiles, and could
be reasonably well fitted. Note the prominent
emission in \Brg. This could not be matched, so we had to concentrate
on the wings. In both cases \HeII2.18 gives an additional constraint
on \Mdot, since at higher values the (theoretical) wings would show too much 
emission.

\paragraph{Cyg\,OB2 $\#$8C and Cyg\,OB2 $\#$8A.} These stars, being of
rather similar type and displaying rather similar profiles (with the
noticeable difference of \HeI1.70, immediately indicating that 8A is
somewhat cooler than 8C), have been carefully analyzed in the optical (and,
again, re-analyzed by R.M.). From the optical, both stars have significantly
different gravities (well constrained from the Balmer line wings), where
object 8C with \g\ = 3.8 has a rather large gravity for its type, cf.
\citet{h02}. The values derived from the IR, on the other hand, are much
closer to each other, namely 3.62 and 3.41, respectively.\footnote{Recently,
\citet{deBecker04} have identified object 8A as 
an O6 I/O5.5 III binary system, therefore the derived
parameters remain doubtful. In our spirit to compare with optical analyses, 
we treated the system as a single star, in accordance with
\citet{h02}.} According to the
observed shape of the profiles and their corresponding theoretical fits, a
higher \g\ would lead to severe inconsistencies. Apart from gravity,
however, the other parameters derived are comparable to their optical
counterparts, including the differences in \Mdot, although the fit quality
of \Brg\ is dissatisfying.

\paragraph{HD\,192639.} For this star, we found a reasonable compromise
concerning the fit quality of the lines present. We derived a \g\ value of
3.3 compared to 3.45 in the optical, because of the wings Br10/11 (note the
different degree of inconsistency in the lines cores!) and due to the shape
of \HeII2.18. With a value of \g\ = 3.45 \HeII2.18 becomes even narrower,
with a more pronounced P-Cygni type profile. The helium abundance was raised
to 0.3 (from 0.2 in the optical) in order to fit the \HeI\ and \HeII\ lines
appropriately in combination with the derived \teff. Also in this case, the
observed \Brg\ line shows a central emission which could by no means be
reproduced. The \HeI2.1607 triplet blend showing up in the theoretical
prediction is not yet present in the observation. 

\paragraph{HD\,210809.} Part of the observed discrepancy in \Brg\ might be
attributed to intrinsic variations in the notoriously variable wind of this
star (\citealt{markova05}), though it is also possible that some (though not
all) of the mismatch arises from errors in the removal of the \Brg\ feature
in the telluric standard. Fortunately, the line wings could be fitted fairly
well, resulting in a mass-loss rate of 5.80 \Mdu\ compared to 5.30~\Mdu\
in the optical. The major difficulty encountered was to fit the \HeI\ and
\HeII\ lines simultaneously. In fact, a decrease in \teff\ leads to an even
more pronounced P-Cygni type profile for \HeII2.18 for the given mass-loss
rate, as was already true for HD\,192639. We regard our solution as the best
compromise possible, accounting for the fact that by a reduction in \teff\
we would also increase the apparent dilemma in Br10/11 and the \HeI\
component in \Brg. The helium abundance was raised by 0.06 to 0.2 in order
to find a compromise for the He lines.

\paragraph{HD\,30614.} For this star a very good fit quality
was obtained making further comments unnecessary.

\paragraph{HD\,209975.} The stellar profiles are fairly well reproduced and
represent the best compromise possible. All hydrogen
features predicted are a little too strong, with some contamination on
the blue side of the profiles. The parameters obtained are comparable 
to the optical ones, except for \g, where we determined 
a smaller value (0.15 dex).

\begin{table*}
\centering

\caption{Adopted and derived stellar and wind parameters obtained from
spectra in the infrared for sample~IV objects {\it not} analyzed in the
optical.  Since no radius information is available, only the optical depth
invariant, $\log Q$, can be derived (Eq.~\ref{defQ}). \teff\ in kK, \Vr\ in
\kms. Centrifugal correction ($\rightarrow \log g_{\rm true}$) assuming a
typical radius.} 
\label{paranew}
\tabcolsep1.5mm
\renewcommand{\arraystretch}{1.1}
\begin{tabular}{l l c c c c c c c}
\hline
Star & Sp.Type & \teff & \g & $\log g_{\rm true}$ & \Y &
\Vr & $\log Q$ & $\beta$ \\
\hline
\hline
HD\,64568  & O3 V((f)) & 45.0 & 3.85 & 3.86 & 0.10 & 150 &- 13.00 & 0.90\\
HD\,46223  & O4 V((f)) & 42.0 & 3.70 & 3.71 & 0.10 & 100 & -12.70 & 0.90\\
HD\,37468  & O9.5 V    & 30.0 & 4.00 & 4.00 & 0.10 &  80 & -14.10  & 1.00\\
\hline
\end{tabular}
\end{table*}

\paragraph{HD\,37128 ($\epsilon$ Ori).} Almost perfect fit. Let us only
point out that the derived value for \teff\ represents an upper limit, since
from this star onwards \HeII\ is no longer present and \HeI\ becomes rather
insensitive to \teff, so that without \HeI2.05 further conclusions are
almost impossible.

\paragraph{HD\,13854 and HD\,13866} have not been analyzed, 
due to missing \HeII\ and \HeI2.05.

\paragraph{HD\,14134.} As above. The ``theoretical'' spectrum displayed
in Fig~\ref{supergiants} shows the insensitivity of the \HeI1.70 and \HeI2.11
lines to \teff\ for temperatures below 30,000~K. Although a virtually perfect
fit has been obtained, the synthetic model (\teff\ = 25,000~K, \g\ = 2.70) 
is located far away from realistic values (roughly at \teff\ = 18,000~K, \g\ = 2.20,
cf. \citealt{kudetal99}).

\section{The ``$\log Q$-approach''}
\label{logqapproach}

Before we compare our results from the IR with optical data, let us briefly
consider those objects where no optical information is available. 
Tab.~\ref{paranew} summarizes the corresponding parameters which constitute
a ``by-product'' of our investigations. Because of the missing radius 
information, we quote the corresponding values for the optical depth
invariant, $\log Q$, instead of the mass-loss rate \Mdot. 

Though we will no longer comment on these stars in the following, we 
would like to point out that all derived parameters appear to be fairly
reasonable, except for the gravity of HD\,46223, which is rather low for a
dwarf of spectral type O4.

It might be questioned how reliable these parameters are, given the fact
that, more precisely, $\log Q$ is actually a scaling-quantity for {\it
recombination lines} formed in the wind. As outlined by \citet{puls05},
however, this quantity is a suitable compromise concerning the scaling
properties of other important physical variables, namely density ($\propto
\Mdote/(\Rstare^2 \vinfe)$) and the optical depth of resonance lines from
major ions (scaling via $\Mdote/(\Rstare \vinfe^2)$). Thus, it might be used
as a {\it general} scaling invariant, though only in an average sense.
Because of these different scaling properties, it is to be expected that
models with identical $\log Q$ but different combinations of \Rstar, \Vi\
and \Mdot\ result in somewhat different profiles, and in the following we
will explore the corresponding uncertainties. By this investigation, we will
also clarify in how far uncertainties in stellar radii (i.e., distances) and
terminal velocities (which will be present in future applications when
analyzing the IR alone, cf. Sect.~\ref{general}) might influence the derived
parameters, particularly $\log Q$. 

\begin{table}
\centering \caption{Different models for $\zeta$~Pup at identical $\log Q =
-12.02$.  The first entry corresponds to our reference model (for additional
parameters, cf. Table~\ref{paraold}). For models A to D we have varied
\Rstar, \Vi\ and \Mdot\ in such a way as to preserve the optical-depth
invariant $Q$ whereas the other parameters remain unchanged. 
\Rstar\ in \Rsuna, \Vi\ in \kms and \Mdot\ in
10$^{-6}$\Msuna/yr (see text and Fig.~\ref{compq}).} \label{tablecompq}
\tabcolsep1.5mm
\begin{tabular}{l c c r c}
\hline
Model & \Rstar & \Vi & \Mdot & $\log Q$ \\
\hline
\hline
HD\,66811  &  19.4 & 2250 & 8.77 & -12.02 \\
HD\,66811-A & 15.0 & 1900 & 4.56 & -12.02 \\
HD\,66811-B & 15.0 & 2550 & 7.19 & -12.02 \\
HD\,66811-C & 25.0 & 1900 & 9.94 & -12.02 \\
HD\,66811-D & 25.0 & 2550 &15.46 & -12.02 \\
\hline
\end{tabular}
\end{table}

\begin{figure}
\resizebox{\hsize}{!} 
{\includegraphics{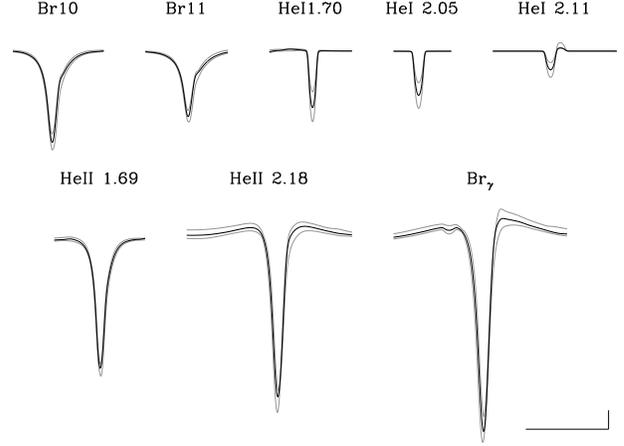}} 
\caption{Variation of synthetic IR spectra for different combinations of
\Rstar, \Vi\ and \Mdot\ but identical optical depth invariant $Q$. Bold:
profiles for our reference model of $\zeta$ Pup. Grey: maximum und minumum
profiles resulting from the models A to D of Table~\ref{tablecompq} and the
reference model, see text. The scale in the lower left corner corresponds to
0.01 microns in wavelength and 1\% of the continuum, respectively. Note that
in almost all cases the maximum/minimum variations are below this value. All
original profiles have been convolved with a typical rotational speed of
$\vsini$ = 100~\kms.} 
\label{compq}
\end{figure}

A first impression of this problem has been given by \citet[Fig.~18]{puls05}
who discussed the variations of the synthetic {\it optical} spectrum of
$\alpha$~Cam when the stellar radius is modified, while keeping the terminal
velocity and $Q$. The differences turned out to be marginal, except for the
innermost cores of a variety of lines. \Ha\ (being {\it the} prototypical
recombination wind line), on the other hand, showed perfect agreement with
the original profile.

We have performed a corresponding analysis in the IR, additionally allowing
for variations in \Vi, by means of our best fitting model for $\zeta$~Pup,
cf. Table~\ref{paraold}. Our choice was motivated by the fact that this
object is one of the few which shows a certain degree of wind-emission in
the wings of \Brg\ and \HeII 2.18, thus displaying profiles which have a
significant contribution from the wind. We have calculated four additional
models, denoted by A to D, where the modified stellar/wind-parameters
(constrained by the requirement of preserving $Q$) are displayed in
Table~\ref{tablecompq}. In particular, we have varied \Rstar\ = 19.4 \Rsun\
by roughly $\pm$ 25\% and \Vi\ = 2250 \kms\ by $\pm$ 300 \kms, consistent
with the typical uncertainties in \Rstar\ and \Vi\ if taken from
calibrations. Within the four models, the mass-loss rates (with a reference
value of 8.5 \Mdu) range from \Mdot\ = 4.6 to 15.5 \Mdu.

In Fig.~\ref{compq} we compare the resulting variations of the corresponding
profiles with our reference solution from Fig.~\ref{supergiants}. Since the
differences turned out to be rather small, we display, at {\it each}
frequency point, the minimum and maximum normalized flux value with respect
to {\it all} five spectra, in order to illustrate the maximum uncertainty
due to the various parameter combinations. As it is evident, in
almost all cases this maximum uncertainty lies well below $\pm$ 1\% of the
continuum flux level, at least if $\vsini$ is not very low. Accounting for
the additional noise within the observation, we would have derived almost
identical stellar parameters including $\log Q$ (with differences well below
the typical errors as discussed in the next section) if a fit by any of the
additional models had been performed. Thus, our hypothesis that $\log Q$ can
be used as a global scaling invariant is justified indeed. 

\newcommand{\rb}[1]{\raisebox{1.5ex}[-1.5ex]{#1}}

\begin{table*}
\begin{center}
\caption{Comparison of stellar and wind parameters in the optical and the
near infrared derived using {\sc fastwind}. \teff\ in kK, \Rstar\ in \Rsuna,
\Mdot\ in 10$^{-6}$\Msuna/yr. \g\ values are corrected for
centrifugal acceleration. If not explicitely indicated, the optical
parameters have been taken from \citet{repo04}. Note that Cyg\,OB2 $\#$8A has 
been recently identified as an O6 I/O5.5 III binary system. In order to compare
with previous analyses, we have retained its single star designation as
used by \citet{h02}}.
\label{paraold}
\tabcolsep1.5mm
\renewcommand{\arraystretch}{1.1}
\begin{tabular}{l l c c l c l c c c l}
\hline
Star & Sp.Type & \Rstar & $T_{\rm eff}^{\rm opt}$ & $T_{\rm eff}^{\rm ir}$ &
$\log g^{\rm opt}_{\rm true}$ &$\log g^{\rm ir}_{\rm true}$ & $Y_{\rm He}^{\rm opt}$ & $Y_{\rm
He}^{\rm ir}$ & ${\dot M}^{\rm opt}$ & ${\dot M}^{\rm ir}$\\ 
\hline
\hline
Cyg\,OB2 $\#$7$^{1)}$& O3 If$^*$ 
                            & 14.6 & 45.5 & 44.0 & 3.71 & 3.71 & 0.21$^{a)}$-0.30 & 0.10 & 9.86 & 10.00 \\ 
HD\,66811  & O4 I(n)f       & 19.4 & 39.0 & 39.0 & 3.59 & 3.59 & 0.20 & 0.17 & 8.80 &  8.77 \\
HD\,15570$^{2)}$& O4 If+    & 22.0 & 42.0 & 38.0 & 3.81 & 3.51 & 0.18 & 0.15 & 17.8 & 15.20 \\
Cyg\,OB2 $\#$8C$^{1)}$& O5 If
                            & 13.3 & 41.0 & 39.0 & 3.81 & 3.62 & 0.09 & 0.10 & 2.25 &  2.00 \\
HD\,14947  & O5 If+         & 16.8 & 37.5 & 37.5 & 3.48 & 3.48 & 0.20 & 0.20 & 8.52 &  7.46 \\
Cyg\,OB2 $\#$8A$^{1)}$& O5.5 I(f) 
                            & 27.9 & 38.5 & 37.0 & 3.51 & 3.41 & 0.10 & 0.10 & 13.5 & 11.50 \\
HD\,192639 & O7 Ib          & 18.7 & 35.0 & 34.0 & 3.47 & 3.32 & 0.20 & 0.30 & 6.32 &  6.32\\
HD\,210809 & O9 Iab         & 21.2 & 31.5 & 32.0 & 3.12 & 3.31 & 0.14 & 0.20 & 5.30 &  5.80 \\
HD\,30614  & O9.5 Ia        & 32.5 & 29.0 & 29.0 & 2.99 & 2.88 & 0.10 & 0.20 & 6.04 &  6.04  \\
HD\,209975 & O9.5 Ib        & 22.9 & 32.0 & 31.0 & 3.22 & 3.07 & 0.10 & 0.10 & 2.15 &  3.30 \\
                  &         &      & 28.5 &      & 3.00 &      & 0.10 &      & 2.40 &       \\
\rb{HD\,37128$^{3)}$}&\rb{B0 Ia}
                            & \rb{35.0} & 27.5 & \rb{29.0$^{b)}$}&  2.95   & \rb{3.01} &0.10& \rb{0.10} &3.01 &\rb{5.25} \\
HD\,15558  & O5 III(f)      & 18.2 & 41.0 & 41.0$^{c)}$ & 3.81 & 3.81 & 0.10 & 0.08 & 5.58 &  7.10 \\
HD\,190864 & O6.5 III       & 12.3 & 37.0 & 36.5 & 3.57 & 3.61 & 0.15 & 0.20 & 1.39 &  0.98 \\
HD\,203064 & O7.5 III       & 15.7 & 34.5 & 34.5 & 3.60 & 3.60 & 0.10 & 0.20 & 1.41 &  2.58 \\
HD\,191423 & O9 III         & 12.9 & 32.5 & 32.0 & 3.60 & 3.56 & 0.20 & 0.20 & $\le$ 0.41 & $\le$ 0.39 \\
HD\,46150$^{4)}$& O5 V((f)) & 13.1 & 43.0 & 40.0 & 3.71 & 3.71 & 0.10 & 0.10 & N/A  &  1.38 \\
HD\,15629  & O5 V((f))      & 12.8 & 40.5 & 40.5 & 3.71 & 3.81 & 0.08 & 0.08 & 1.28 &  1.28 \\
HD\,5689$^{2)}$& O6 V       &  7.7 & 37.0 & 36.0 & 3.57 & 3.66 & 0.33 & 0.20 & 0.16 & 0.17 \\
HD\,217086 & O7 Vn          &  8.6 & 36.0 & 36.0 & 3.72 & 3.78 & 0.15 & 0.15 & $\le$ 0.23 & $\le$ 0.09 \\
HD\,13268  & ON8 V          & 10.3 & 33.0 & 33.0 & 3.48 & 3.48 & 0.25 & 0.25 & $\le$ 0.26 & $\le$ 0.17 \\ 
HD\,149757 & O9 V           &  8.9 & 32.0 & 33.5 & 3.85 & 3.85 & 0.17 & 0.17 & $\le$ 0.18 & $\le$ 0.15 \\
           &                &      & 31.4 &      & 4.24 &      & 0.10 &      & 0.009 &    \\
\rb{HD\,149438$^{5)}$}&\rb{B0.2 V}&\rb{5.3}&31.9&\rb{31.0}&4.15&\rb{4.00}&0.12&\rb{0.10}&0.02{\ldots}0.06&\rb{0.020}\\
\hline
\end{tabular}
\end{center}
\smallskip
Optical parameters taken from\\
$^{1)}$ \citet{h02} \quad
$^{2)}$ \citet{h00} (unblanketed {\sc fastwind} models)\\
$^{3)}$ \citet[upper entries]{kudetal99} and from \citet[lower entries, 
\Mdot\ scaled to \Rstar\ = 35 \Rsuna]{urban04} \\   
$^{4)}$ \citet{h92} (unblanketed plane-parallel H/He models) \\
$^{5)}$ \citet{Kilian91} and from \citet{PB04} with respect to wind
properties (upper entries)\\ \hspace*{2ex} and from R.M. ({\sc fastwind},
lower entries); the limits of \Mdot\ correspond to velocity field exponents
$\beta = 2.4{\ldots} 0.8$.\\
$^{a)}$ from a re-analysis by R.M.({\sc fastwind})\\ 
$^{b)}$ upper limit \\
$^{c)}$ taken from optical analysis\\ 
\end{table*}

By comparing the individual spectra from the four models with our reference
spectrum, it turned out that the major source of disagreement results from
models A and D, whereas models B and C show almost identical profiles. The
reason for this discrepancy/agreement is readily explained by noting that
the run of $\rho$ as a function of $\taur$ (which most importantly controls
the emergent spectrum) is very similar for the reference model and models B
and C, whereas for model A higher densities and for model D lower densities
(as a function of $\taur$) are found, particularly in the transition region
between photosphere and wind. Consequently, the recombination rates in model
A are somewhat larger, leading to deeper \HeI\ profiles, and vice versa for
model D. The resulting uncertainty with respect to \teff, however, lies 
well below $\pm$ 500~K. 

Note that all these investigations have been performed by keeping $\beta$ at
its reference value. If $\beta$ cannot be constrained from the emission
wings of wind lines, the uncertainty in $\log Q$ can become severe,
extending to a factor of two (cf. \citealt{puls96}).

\section{Comparison with optical data}
\label{compdata}

In the following, we now return to the most important question, to what extent
a lone near IR analysis is able to recover the parameters from an
analogous optical analysis. The corresponding data can be found in
Tab.~\ref{paraold}.

\begin{figure*}
\begin{minipage}{8.8cm}
\resizebox{\hsize}{!}
   {\includegraphics{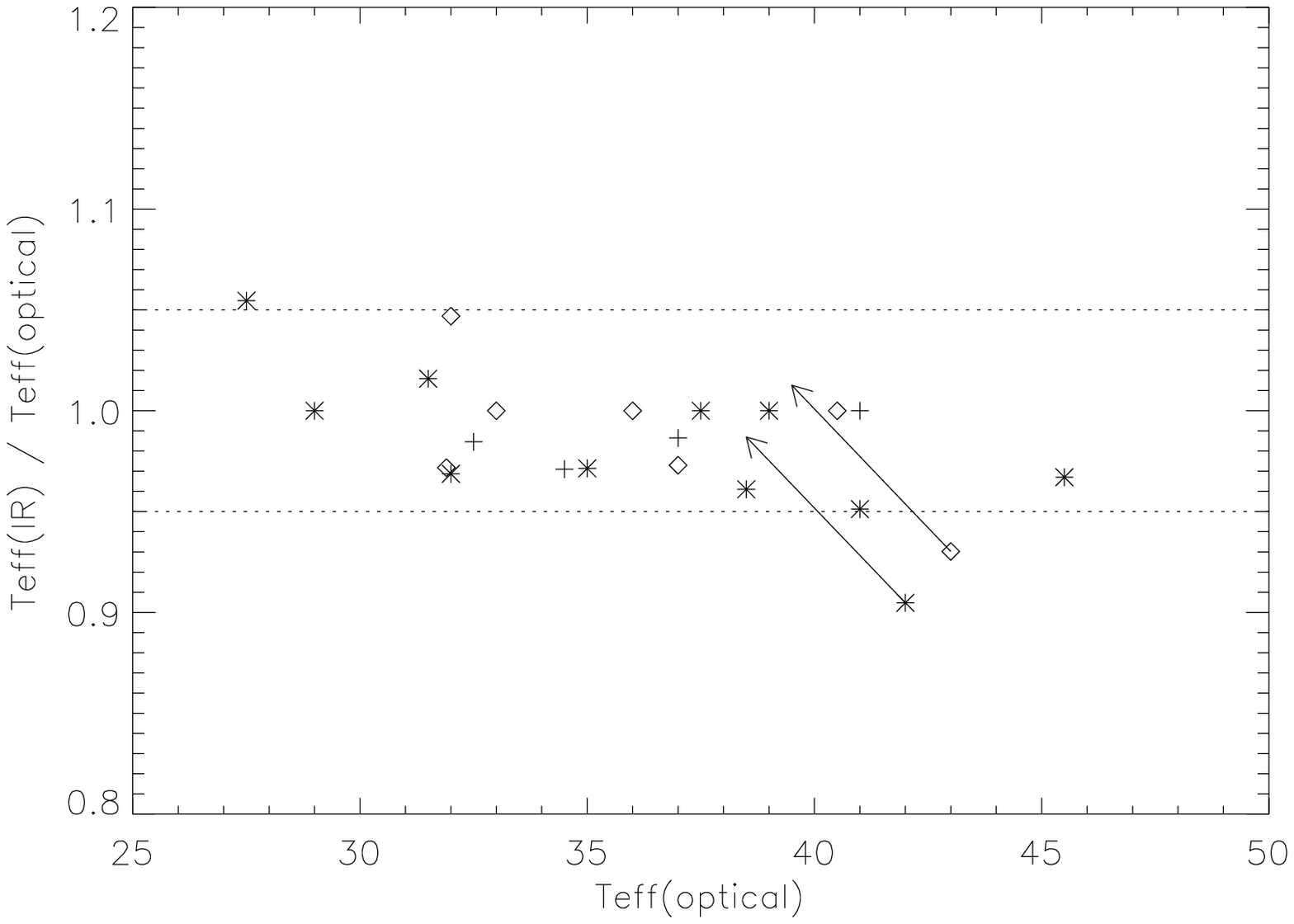}}
\end{minipage}
\hfill
\begin{minipage}{8.8cm}
   \resizebox{\hsize}{!}
      {\includegraphics{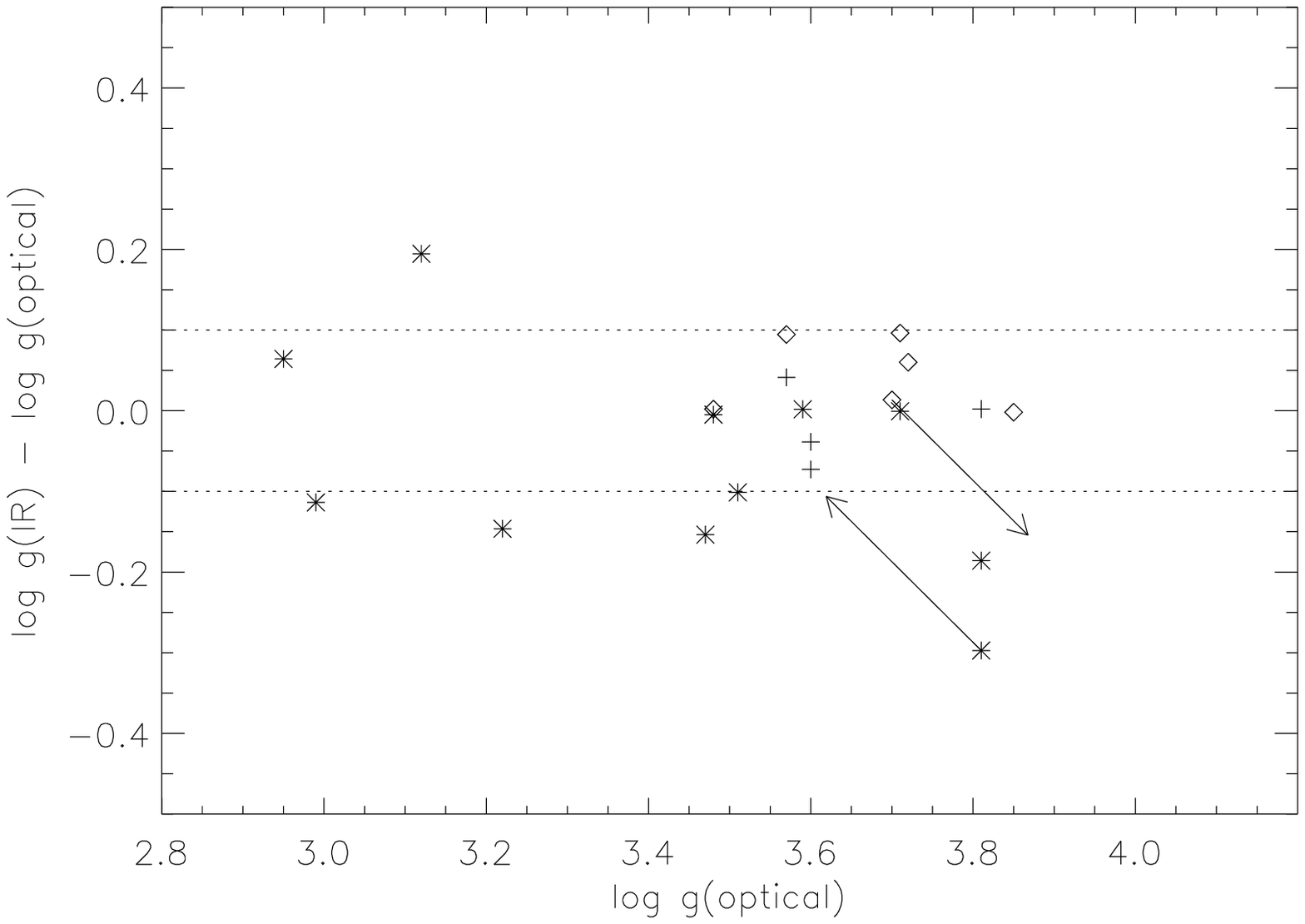}}
\end{minipage}
\caption{Comparison of \teff\ and \g\ derived from the optical and the
near-IR. Asterisks, crosses and diamonds correspond to luminosity classes I,
III and V, respectively. The displayed ``error bars'' correspond to maximum
uncertainties claimed in optical analyses, namely $\pm$5\% in \teff\ and
$\pm$0.1 in \g. The two arrows in the left panel correspond to the objects
HD\,15570 and HD\,46150 (analyzed by means of unblanketed models in the
optical) and indicate the average shift in position if blanketing effects
would have been accounted for. The appropriate shifts in \g, applying the
\g\ - \teff\ calibrations from \citet{repo04}, have been indicated on the right,
again by arrows. $\tau$ Sco has not been included into the right figure,
since the IR gravity could not be constrained due to missing Br10/11.
Concerning the remaining outliers, see Sect.~\ref{compdata}.} 
\label{compopt1}
\end{figure*}

\begin{figure}
\resizebox{\hsize}{!}
   {\includegraphics{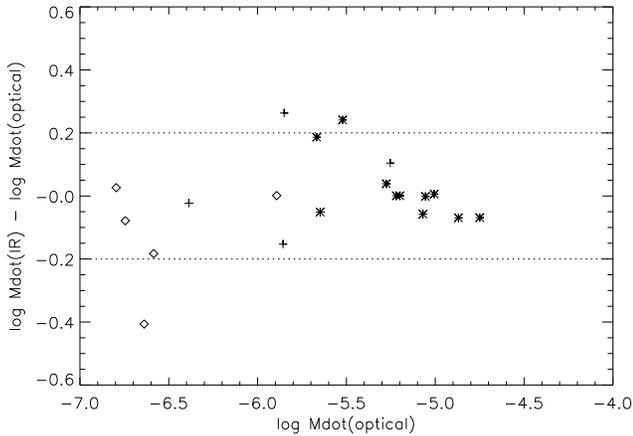}}
\caption{As Fig.~\ref{compopt1}, but for mass-loss rate \Mdot. Upper
limits have been treated at their nominal values. The ``error bars''
correspond to mean uncertainties of $\pm$0.2 dex quoted 
for \Ha\ measurements. $\tau$ Sco has not been included in this comparison
(see text).}
\label{compmdot}
\end{figure}

Figs.~\ref{compopt1} and \ref{compmdot} compare the results for \teff, \g\
and \Mdot\ for all stars which have been analyzed in the optical. In each
figure, we have indicated error bars usually quoted for the corresponding
``optical'' quantity. In particular, the {\it maximum} errors for effective
temperature are on the order of $\pm$5\% (corresponding to $\pm 2,000~K$ at
\teff\ = 40,000~K\footnote{This uncertainty is also consistent with the
uncertainty related to the \HeI\ singlet problem possibly affecting our
optical analyses, cf.~\citet[ Sect.~10]{puls05}.}) and {\it typical} errors
for \g\ are $\pm 0.1$. The ``error bars'' for the mass-loss rates, indicated
as $\pm$0.2 dex, correspond to {\it mean} values attributed to \Mdot\
measurements from \Ha. Note, however, that the actual precision is an
increasing function of \Mdot, being higher than 0.2 dex for low \Mdot\ and
lower for larger values (e.g., \citealt{puls96}). Remember also that all our
simulations (both in the IR and the optical) have been performed with {\it
un-clumped} models, i.e., the derived mass-loss rates represent upper limits
and may need to be corrected.

From the three figures, it can immediately be seen that the majority of 
IR-values are in reasonable agreement with the corresponding optical data.
Most importantly, no {\it obvious} trend is visible, neither as a function
of the parameter itself nor as a function of luminosity class (``lc'').
A weak trend in \teff\ cannot be excluded though: From
\teff\ = 35,000~K on, the IR data are distributed more towards lower values
(than derived from the optical).

In the following, we will briefly discuss the outliers, i.e., those objects
which are located beyond the indicated error bars and thus must be interpreted
as severe mismatches. 

With respect to \teff, three objects seem to be discrepant, at least at
first glance. Concerning the moderate deviation of HD\,37128
($\epsilon$~Ori, with an optical \teff\ = 27,500~K), remember that our
IR-value is an upper limit only. In particular, we have compared our results
with those derived from a recent, line-blanketed re-analysis by
\citet{urban04}, exploiting the ionization balance of silicon (lower entries
in Tab.~\ref{paraold}). The other two outliers, HD\,46150 (lc~V, appearing
to be 3,000~K cooler) and HD\,15570 (lc~I, 4,000~K cooler), are
unproblematic as well, since the corresponding optical analyses have been
performed by {\it unblanketed} models, and the obtained differences are
just of the expected order of blanketing effects at \teff\ = 42,000~K, i.e.,
roughly $\Delta \Teffe \approx -3,500$~K (e.g., \citealt{m02, repo04}). New
positions corrected for blanketing effects of this amount have been
indicated by arrows. 

Blanketing effects do not only affect effective temperatures, but also
gravities. In particular, one must still be able to fit the Balmer line
wings, which, for the correspondingly cooler temperatures, usually
yields lower values for \g. Because of possible differences in the derived
He-abundance, however, the actual correction depends strongly on the
specific situation, and corrections towards significantly lower values and
of negligible amount have been found in parallel, see
\citet[Fig.~16]{repo04}.

In order to obtain at least an impression about the situation, we have used
the \g\ - \teff\ calibrations provided by Repolust et al. (their Fig.~17, see
also \citealt{markova04}). These calibrations have been derived from their
analysis of Galactic O-stars, utilizing the same code as applied here
(augmented by the results for O-dwarfs from \citealt{m02}), and should be 
valid within $\pm 0.1$ dex, at least differentially. Using the corrected
values for \teff\ from above (i.e, 3,500~K cooler than the unblanketed ones), 
we find \g\ = 3.62 and 3.87 for HD\,15570 and HD\,46150, respectively, where
the former value could be even lower, due to the extreme character of the
object (O4If+). The corrected positions have been indicated in
Fig.~\ref{compopt1}, right panel, again by arrows. Obviously, the IR-gravity
of HD\,15550 is now almost consistent with the calibrated value from the
optical. For HD\,46150, on the other hand, the situation has become worse,
with a discrepancy of -0.15 dex, {\it if} our calibration applies for this
star. Note, however, that also the unblanketed optical gravity is (very) low
for a dwarf of this spectral type, which might indicate that this object is
somewhat peculiar, and an optical re-analysis is certainly required to obtain
firm conclusions.

$\tau$ Sco has not been included into this comparison, since no
photospheric hydrogen lines have been observed and its gravity is almost
unconstrained with respect to our NIR analysis (cf.
Sect.~\ref{dwarfcomments}). Remember that we have favoured the lower value
alone because of the shape of the central emission in \Brg, which
constitutes the only difference between a \g\ = 4.0 and \g\ = 4.25 model,
given the observed lines. In future analyses with only IR spectra available
we would favor this fit under the same circumstances anyway.

Thus, in total we have four definite outliers (i.e., above the
1-$\sigma$ level and discarding HD\,46150 because of its unclear
status). HD\,192639 and HD\,209975 appear to be lower in
gravity by -0.15 dex when compared to the optical and the record holders,
Cyg\,OB2 \#8C and HD\,210809, give differences of -0.2 and +0.2 dex,
respectively.  In this sample, we might also include HD\,46223 from
Tab.~\ref{paranew}, since the derived gravity is probably too low by a
similar amount. 

Concerning mass-loss rates, the situation is as satisfactory as described
above. First note that Fig.~\ref{compmdot} also displays those stars for
which we can only provide upper limits of \Mdot, and which we have compared.
The only star missing in this comparison is $\tau$~Sco, however a comparison
with both the ``optical'' mass-loss rate and the value cited in
Tab.~\ref{paraold} (which has been derived from an alternative IR analysis,
cf. Sect.~\ref{dwarfcomments}) reveals a disagreement of a factor of roughly two
(smaller and larger, respectively). Such a difference is not too bad, taken
the intrinsic uncertainties at such low wind densities. We will come back to
this problem later on.

As expected from the non-linear increase of wind-emission as a function of
\Mdot, the disagreement between optical and near IR mass-loss rates becomes 
smaller for larger wind-densities. For $\log \Mdote^{\rm opt} \ga -5.3$,
these differences are lower than 0.1 dex, which indicates the sensitivity of
the \Brg\ line wings (remember that only the wings could be fitted in high
\Mdot\ objects) and partly of \HeII\ on this parameter. The differences
obtained for the corresponding equivalent widths (observations vs. theory,
cf. \citealt {len04} and Sect.~\ref{fitstrat}) are thus almost exclusively
due to the differences with respect to the line cores, which at present 
cannot be explained conclusively, though a relation to wind-clumping might
be possible (cf. Sect.~\ref{summary}). For the low \Mdot\ stars, on the
other hand, no trend (particularly not towards considerably larger \Mdot) 
is visible.  The above problem might be related to the physical
conditions in the ``intermediate'' wind, whereas close to the wind base no
obvious difference between the formation of \Ha\ and \Brg\ seems to be
present.

This hypothesis is consistent with the fact that the largest deviations
are found at intermediate wind-strengths, i.e., in those cases where \Brg\
is considerably refilled but does not show emission wings. The
most prominent difference, $\Delta \log \Mdote$ = 0.33, is found for
HD\,37128, if we compare with results from the unblanketed analysis by
\citet{kudetal99}. The corresponding re-analysis by means of blanketed
models (\citealt{urban04}, see above) yields a moderately larger mass-loss
rate\footnote{though at a lower effective temperature and a somewhat larger
value of $\beta$, which, in combination, are canceling each other with respect
to \Mdot.}, and the remaining discrepancy, $\Delta \log \Mdote$ = 0.25 has been
indicated in Fig.~\ref{compmdot}. The difference for the other outlier,
HD\,203064, is of the same order, $\Delta \log \Mdote$ = 0.26, and has to be
considered as ``real'' as well, since all other parameters do agree
(except for \Y, which has only marginal influence on the derived mass-loss
rate). For HD\,217086, finally, differences occur only with respect to an
upper limit, e.g., the IR analysis predicts a lower limit than the optical
one.

Concerning the helium abundance (not plotted), there are only two
problematic cases. Cyg\,OB2 \#7 has been discussed already in
Sect.~\ref{sgcomments}, and both the optical abundance (\Y\ = 0.2 {\ldots}
0.3) and the corresponding IR value (\Y\ = 0.1 ) are uncertain, the latter
due to the missing \HeI\ lines (although we would derive a significantly
lower \teff\ if the high abundance were true). The second outlier is
HD\,5689 with \Y\ (optical) = 0.33 and \Y\ (IR) = 0.20. Again, this
discrepancy is probably irrelevant, since the optical analysis has been
performed by means of unblanketed atmospheres, which are well-known to
overestimate the helium abundance in a number of cases (cf. \citealt{repo04},
particularly Sect.~7.2).  

\begin{figure*}
\resizebox{\hsize}{!} {\includegraphics{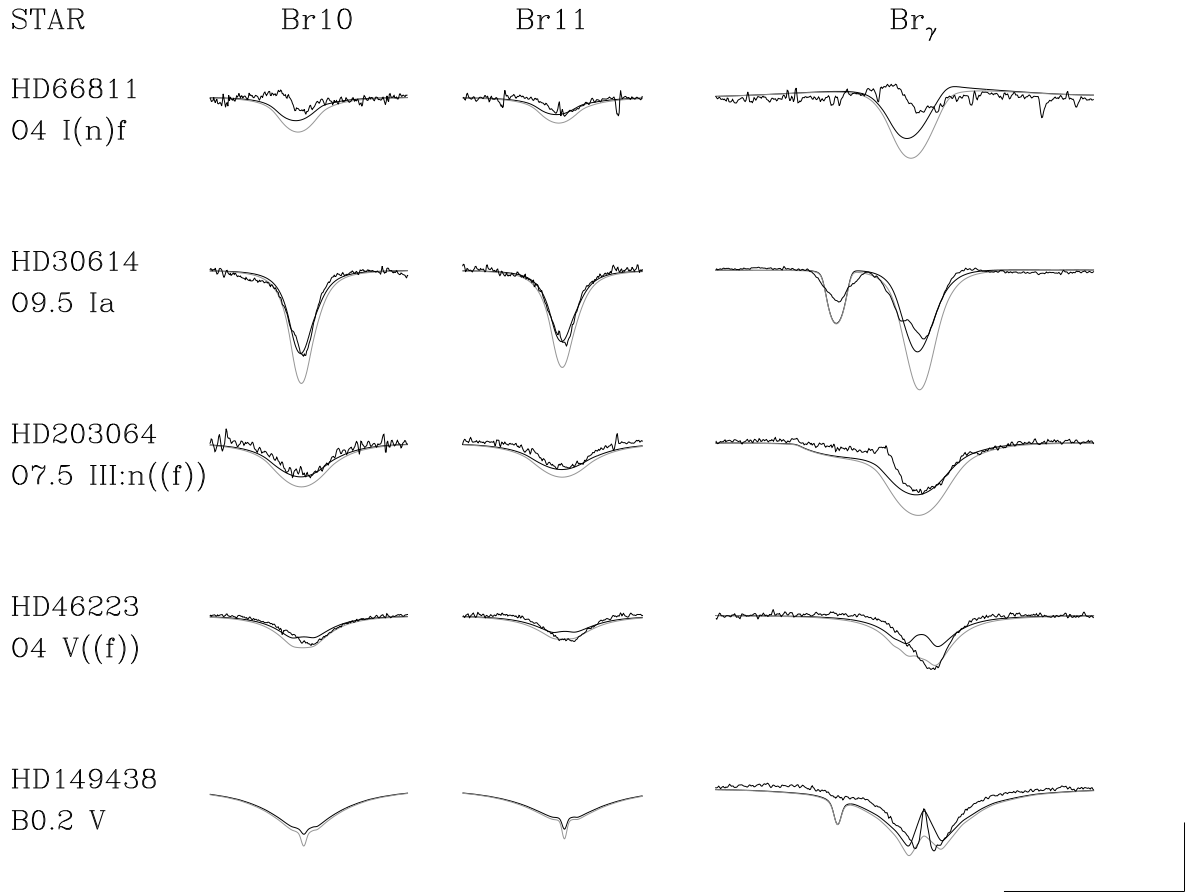}}
\caption{Influence of different collisional cross sections for hydrogen:
some prototypical examples. Solid: Best fit with cross sections from
\citet{gio87}, as in Figs.~\ref{dwarfs} to \ref{supergiants}. Grey: Models
with {\it identical parameters}, but cross sections calculated according
to \citet{PB04}. See text.} 
\label{compcoll}
\end{figure*}

\subsection{Comments on hydrogen collisional cross sections}
\label{collhyd}

In Sect.~\ref{atomdat} we briefly discussed the importance of
consistent collisional data for the resulting IR line profiles. We  
outlined recent calculations performed by \citet{PB04} and references
therein. These authors provide a recipe for an ``optimum'' choice of
collisional data, based on a number of comparisons with observations,
comprising mostly BA-type dwarfs (including $\tau$~Sco) and supergiants,
whereas only one test has been presented for an O-star, the O3.5 {\it dwarf}
HD\,93250. 

After incorporating their data into our version of {\sc fastwind}\footnote{
Note that Przybilla \& Butler have used a similar version to check part of 
their calculations.}, we have subsequently tried to analyze our observed
dataset. First let us mention that this modification gave rise to
changes in the hydrogen lines alone, since in all cases the temperature
structure (being dependent on the collisional bound-bound rates) remained
almost unaffected, with maximum changes on the order of 100~Kelvin.

Unfortunately, however, it turned also out that, again in {\it all cases},
the hydrogen line cores became stronger (in agreement with the findings by
Przybilla \& Butler), whereas the line wings are barely affected, as shown
in Fig.~\ref{compcoll} for some prototypical examples taken from the fits in
Figs.~\ref{dwarfs} to \ref{supergiants}. In a few cases this might actually
lead to an improvement of the situation, e.g., for HD\,46223 or Cyg\,OB2
\#8C, which actually need gravities higher than those derived above.
(Remember that increasing the gravity results in shallower line cores, cf.
Sect.~\ref{stark}). Since, however, the IR-gravities based on our standard
collisional data from \citet{gio87} were found to be consistent with the
optical ones in the {\it majority} of cases, models based on the alternative
data by Przybilla \& Butler would consequently lead to an overestimate of
gravities.

The same would be true for the mass-loss rates. Using the new data would 
sometimes improve the situation, e.g., any central emission inside \Brg\ (if
present) becomes reduced, cf. HD\,46233 and $\tau$ Sco in
Fig.~\ref{compcoll}. Actually, this reduction is the origin of the lower
mass-loss rate of $\tau$~ Sco as derived by \citet{PB04}: If \Mdot\ is
decreased, the strength of this feature increases again, at least if the
winds are very weak, due to subtle NLTE effects and unrelated to any
direct wind emission. 

For ``normal'' winds, on the other hand, where wind emission plays the
primary role, the deeper cores predicted by the ``new'' models would
necessitate higher mass-loss rates. Thus, the present situation would get
worse, at least in those cases where \Mdot\ is no longer derived exclusively
from the wings. As an example, consider HD\,203064 in Fig.~\ref{compcoll},
which presently has an IR mass-loss rate which is a factor of two larger
already. Using the new dataset would further deteriorate this discrepancy.  

Insofar, the preference for our standard set of collisional data is
triggered solely from the results described above, namely from the generally 
satisfactory agreement between the IR and optical analyses {\it for those
objects analyzed in the present investigation}. We do not argue that one set
or the other is better, but point out that in these cases our
standard dataset gives results which are more consistent with the optical. 

Of course, we have also looked into some of the details responsible for the
differences obtained. It turned out that the NLTE departure coefficients are
astonishingly similar, when comparing the results from both collisional
datasets. There are only (very) subtle differences in those regions where
the line cores are formed giving rise to the deeper profiles if the dataset
by Przybilla \& Butler is used. Either the lower level of the transition is
slightly more populated, or the upper one is slightly less populated. The
obvious discrepancies are then induced by the extreme sensitivity of the IR
line formation on such subtle differences. As has already been argued about
the formation of \HeI\ lines (Sect.~\ref{nlteeffects}), this discrepancy
would barely be visible if the lines were situated in the optical. Insofar,
not only the data but also the numerical treatment plays a crucial role. In
any case it is quite astonishing how well the observed profiles can be
simulated.

\section{Summary and conclusions}
\label{summary}

In this paper, we have analyzed 25 Galactic O and early B-stars by means of 
{\it H} and {\it K} band spectroscopy. The primary goal of this
investigation was to check to what extent a lone near-IR spectroscopy is
able to recover stellar and wind parameters derived in the optical. This is
critical to our desire to precisely analyze the hot, massive stars, deep 
within the disk of our Galaxy, and in particular the very young, massive 
stars just emanating from their birth places.

Most of the spectra have been taken with {\sc subaru-ircs}, at an
intermediate resolution of 12,000. In order to synthesize the strategic H/He
lines present in the {\it H/K} band, we have used our recent, line-blanketed
version of {\sc fastwind}. In total, seven lines have been investigated, three
from hydrogen, including \Brg\ serving as a diagnostic tool to derive
wind-densities, two \HeI\ and two \HeII\ lines. For two stars, we could
make additional use of \HeI2.05 (singlet) which has been observed with {\sc
irtf-cshell}. Apart from \Brg\ and \HeII2.18, the other lines are predominately
formed in the stellar photosphere, and thus remain fairly uncontaminated
from more complex physical processes, particularly clumping.

In our attempt to prepare all required broadening functions, it turned out 
that at present we have to rely on the Griem approximation for 
Stark broadening (important for hydrogen and \HeII ), since the corresponding
published data (based on the more exact VCS approach) suffer from 
numerical problems, particularly for the members of higher series.

First we investigated the predicted behaviour of the strategic lines, by
means of a large model grid described in \citet{puls05}. Interestingly and
in contradiction to what one expects from the optical, almost all photospheric
lines in the {\it H} and {\it K} band  
(from H, \HeI\ and \HeII\ ) {\it become stronger if the gravity decreases}. In
Sect.~\ref{predictions}, we have carefully investigated the origin of this
rather unexpected behaviour.

Concerning H and \HeII, it is related to the particular behaviour of
Stark broadening as a function of electron density, which in the line cores
is somewhat different for members of lower and higher series. For the
latter, the cores become deeper when the density decreases, and
contribute more to the total line strength than in the optical. 

Regarding \HeI, on the other hand, the predicted behaviour is due to some
subtle NLTE effects resulting in a stronger overpopulation of the lower
level when the gravity decreases, so that the source function becomes weaker
and the profile deeper, i.e., stronger. This strong dependence of the
profile on the source function is a direct consequence of the IR line
formation with $h \nu/kT << 1$.  If those lines were situated in the
optical, on the other hand, optical depth effects would dominate, leading to
a decrease of line strength due to a lower number of \HeI\ absorbers. This
explains the different (and ``normal'') behaviour of, e.g., \HeI4471.

In Sect.~\ref{comp}, we have compared our calculations with results 
presented recently by \citet{len04}, utilizing the alternative NLTE model
atmosphere code {\sc cmfgen}. In most cases, we found reasonable and
partly perfect agreement. Only the \HeI2.05 singlet for mid O-types 
suffers from some discrepancy, in agreement with our analogous findings
for optical \HeI\ singlets (\citealt{puls05}).

After carrying out the analysis for our sample described above (and in
agreement with the predictions from our model grid), we find that an {\it
H/K} band analysis is able to derive constraints on the same set of stellar
and wind parameters as it is known from the optical, e.g., \teff, \g, \Y\
and optical depth invariant $Q$, where the latter yields the mass-loss rate
\Mdot\ if stellar radius and terminal velocity are known. For cooler
objects, when \HeII\ is missing, a similar analysis might be possible if
\HeI2.05 (provided to be described correctly) is available and
the helium content can be adopted (due to the almost orthogonal reaction 
of \HeI2.05 and \HeI2.11 on \teff\ and \g). This should be possible for 
very young objects containing unprocessed material.

For future purposes, when no UV observations will be available, the terminal
velocity \Vi\ has to be taken from calibrations, as it is true for the
velocity field exponent $\beta$, at least in those cases when no emission
wings in \Brg\ are visible. Concerning the determination of \Rstar, a similar
strategy as in the optical might be developed, utilizing infrared colors and
distances.

For most of our objects, we obtained good fits, except for the line cores of
\Brg\ in early O-stars with significant mass-loss (see below), and except
for the fact that particularly at mid O-types Br10/11 could not be fitted in
parallel. We have argued that this discrepancy is similar to the problem in
the optical, concerning \HeII4200/4541 (e.g., \citealt{h02}). Due to the
similarity in the involved levels and broadening functions, we have
speculated about a possible defect of these broadening functions for
transitions between members of higher series. The largest discrepancy,
however, was found for the line cores of \Brg. First note that this problem
is not particularly related to our code, since also {\sc cmfgen} exhibits the
same shortcoming. Whereas the observations show \Brg\ mostly as rather
symmetric emission lines, the models predict a P~Cygni type line, with a
comparably deep core which is never observed. Note that this type of profile
can only be created if the {\it ratio} of departure coefficients for the
involved levels (n = 4 \rarrow 7) deviates strongly from unity (cf.
\citealt{puls96}), whereas a ratio close to unity would just give the
observed symmetric emission profile. One might speculate that this can be
achieved due to a stronger influence of collisional bound-bound processes,
which, e.g., might be possible in a strongly clumped medium. Remember,
however, that \Brg\ typically forms inside the \Ha-''sphere''
(\citealt{len04}), where the degree of clumping is usually thought to be
moderate (e.g., \citealt{markova04, repo04}), though a recent investigation
by \citet{bouret05} strongly indicates the opposite.

After having derived the stellar and wind parameters from the IR, we have
compared them to results from previous optical analyses, in an almost
strictly differential way, since most of these results have been obtained
also on the basis of {\sc fastwind}. Overall, the IR results coincide in
most cases with the optical ones within the typical errors usually quoted
for the corresponding parameters, i.e, an uncertainty in \teff\ of 5\%, in
\g\ of 0.1 dex and in \Mdot\ of 0.2 dex, with lower errors at higher wind
densities. In most of the cases where we have found discrepancies beyond
these errors, their origin could be easily identified. {\it Definite}
outliers above the 1-$\sigma$ level where found in four cases with respect
to \g\ and in two cases for \Mdot, at intermediate wind-strengths.
Given the 1-$\sigma$ character of the discussed errors, these mismatches are
still not too worrisome, since all identified discrepancies lie well below
2-$\sigma$, and no trends are visible. Particularly with respect to \g\, it
is quite possible that errors exceeding the nominal error of 0.1 dex do
exist in certain cases, both in the optical and the IR, related to fits
which are not at their (global) optimum (e.g., \citealt{mokiem05}).
Re-assuring is the fact that the mean difference between NIR and
optical gravities is very close to zero, $\Delta \log g = -0.025 \pm 0.1$.

As a by-product of our investigation, we could determine the (IR-) stellar
parameters and the $\log Q$ value for three dwarfs, which have not been
analyzed in the optical so far. Though we have shown that the derived 
quantities are rather insensitive to uncertainties in \Rstar\ and \Vi,
these objects need to be checked in the optical.

Let us highlight one additional ``bonus'' obtained from the
infrared. In those cases when a star has an extremely weak wind and the
core of \Brg\ can be resolved (requiring a very low rotational speed), the
central emission will give us a clue about the actual mass-loss rate and not
only an upper limit, as is true for the optical. An example of this kind
of diagnostics is $\tau$~Sco. Particularly with
respect to recent investigations of young dwarfs with surprisingly weak
winds (\citealt{m04}), this will turn out as an invaluable source of
information (even more, if coupled with observations of \Bra, e.g.
\citealt{naj98, PB04}).

After finishing this investigation, we are now able to constrain the
observational requirements to perform such a {\it detailed} IR-analysis (but
see below). Most important is a high S/N, because most of the lines to be
investigated are extremely shallow, at least for the hotter objects. The 
resolution must be sufficient to disentangle the line
cores from the wings (particularly important for \Brg) and to obtain
reasonable clues about any contamination due to reduction problems. As for
the required set of lines, almost all lines analyzed in the present paper
are necessary to obtain useful constraints, maybe except for Br10, since
Br11 seems to be less contaminated. Since both \HeII\ lines behave very
similarly and show the same degree of consistency or disagreement (if
present), one of those two lines might be discarded as well.

In the last section of this paper, we have argued that our standard
implementation of hydrogen collisional cross sections seems to give results
which are in better agreement with the optical results for our sample of
{\it hot} objects, compared to the data suggested recently by \citet{PB04}.
In view of their findings, namely that for cooler stars their prescription
gives more consistent results, this discrepancy has to be clarified in
future work. The same of course is true regarding the severe mismatch of the
\Brg\ cores. A first step will require to include clumping and to
investigate to what extent this process might improve the situation.

The value of a reliable quantitative analysis for hot, massive stars based
entirely in the infrared cannot be overstated. Most obvious, it will allow
the evaluation of massive star characteristics at an evolutionary stage
significantly earlier than has ever been possible before. The influence of
disk emission may render the photosphere of some very young massive stars
inaccessible. We suspect, however, that among the most massive stars, around
mid-O or hotter, the disk will be destroyed well before even near-infrared
studies would be feasible due to the very short disk lifetime 
(\citealt{wathan97}).

A combined {\it H} and {\it K} band analysis with the S/N we present
here won't always be possible, nor is it required. A S/N of just 150 and
resolution of order $R \sim 5000$ will typically be sufficient for a
reasonable analysis, provided the star is not a slow rotator (see discussion
in \citealt{hanson05}). How deep into the Galaxy one can probe is a strong
function of the line of sight extinction and the intrinsic stellar
brightness. For late-O dwarfs with $A_{\rm V} \sim 20$, one can only probe a
few kiloparsecs. However, for early-O dwarfs or most O and early-B
supergiants, even with $A_V \sim 20$, one should be able to probe the entire
solar circle. For this level of extinction, it is the {\it H} band
measurements which pose the greatest challenge, though $S/N \sim 120-150$
should still be attainable with an 8-m class telescope. For still redder
stars, the {\it H} band would be too dim for useful spectra. Once $m_K >
14-15$ (OB supergiants at the far side of the Galaxy with $A_{\rm V} \sim
30$), the corresponding low S/N for such a dim star wouldn't allow a robust
quantitative analysis, though spectral classification is still possible.
Thus, excluding sight-lines where the extinction has become extreme (A$_{\rm
V}$ $>$ 30), nearly every luminous OB star within our Galaxy becomes
accessible through near-infrared spectroscopic studies.

One of the powers of quantitative analysis is its ability to
determine absolute magnitudes. When OB stars are in clusters, those cluster
distances will be robust, giving us clues to the structure and nature of the
presently, poorly understood, inner Milky Way.  OB stars serve as
secondaries to massive compact objects.  Because the extinction is typically
high for such systems found in the inner Galaxy, a NIR analysis of the OB
companion provides the only means for making critical measurements to
constrain these fascinating systems. In truth, because many more O
stars within our Galaxy are visible in the NIR than the optical {\it by
almost two orders of magnitude}, the development of a robust quantitative
analysis in the infrared will stimulate entirely new, important results on
massive stars, their formation and evolution and numerous valuable insights
into the inner workings of our Milky Way Galaxy.

\acknowledgements{We like to thank our referee, Dr. Pat Morris, for useful
comments and suggestions, and Dr. Keith Butler for stimulating
discussions and assistance. Many thanks also to Dr. Paco Najarro for ongoing
discussions concerning subtle {\sc nlte}-effects.
We are grateful to Dr. Alan Tokunaga for his
expertise and guidance in the use of the Subaru and IRTF
telescopes.

T.R. appreciates financial support in the form of a grant by the International
Max-Planck Research School on Astrophysics (IMPRS), Garching.
M.M.H. was supported by the National Science Foundation under Grant No.
0094050 to the University of Cincinnati, and J.P. gratefully acknowledges
support by NATO Collaborative Linkage Grant No. PST/CLG 980007.

\end{document}